\documentclass[aps,pre,preprint,groupedaddress,showpacs]{revtex4-1}
\usepackage[dvips]{graphicx}
\usepackage{textcomp}
\usepackage{amssymb}

\newcommand{\mc}{\multicolumn}
\begin{document}

\title{
\Large\bf Monte Carlo study of a generalized icosahedral model on the simple 
  cubic lattice}

\author{Martin Hasenbusch}
\affiliation{
Institut f\"ur Theoretische Physik, Universit\"at Heidelberg,
Philosophenweg 19, 69120 Heidelberg, Germany}
\date{\today}

\begin{abstract}
We study the critical behavior of a generalized icosahedral model on the 
simple cubic lattice. The field variable of the icosahedral model might
take one of twelve vectors of unit length, which are given by the normalized
vertices of the icosahedron, as value. Similar to the Blume-Capel model, where 
in addition to $-1$ and $1$, as in the Ising model, the spin might take the  
value $0$, we add in the generalized model $(0,0,0)$ as allowed value. There is 
a parameter $D$ that controls the density of these voids. For a certain range  
of $D$, the model undergoes a second-order phase transition. On the critical 
line, $O(3)$ symmetry emerges. Furthermore, we demonstrate that within this
range, similar to the Blume-Capel model on the simple cubic lattice, there is a
value of $D$, where leading corrections to scaling vanish.
We perform Monte Carlo simulations for lattices of a linear size up to $L=400$ 
by using a hybrid of local Metropolis and cluster updates. The motivation
to study this particular model is mainly of technical nature. Less 
memory and CPU time are needed than for a model with $O(3)$ symmetry at
the microscopic level. As the result of a finite-size scaling analysis we
obtain $\nu=0.71164(10)$, $\eta=0.03784(5)$, and $\omega=0.759(2)$ for 
the critical exponents of the three-dimensional Heisenberg universality class.
The estimate of the irrelevant renormalization group eigenvalue 
that is related with the breaking the $O(3)$ symmetry is $y_{ico}=-2.19(2)$.
\end{abstract}

\maketitle
\section{Introduction}
In the neighborhood of a second-order phase transition, thermodynamic 
quantities diverge, following power laws.
For example, the correlation length behaves as
\begin{equation}
\label{xipower}
\xi = a_{\pm} |t|^{-\nu}  \; \left(1+b_{\pm} |t|^{\theta} + c t + ...\right)
\;,
\end{equation}
where $t=(T-T_c)/T_c$ is the reduced temperature. The subscript $\pm$ of the
amplitudes $a_{\pm}$ and $b_{\pm}$ indicates the high ($+$) and the low
($-$) temperature phase, respectively. There are non-analytic or confluent
and analytic corrections. The leading ones are explicitly given in 
eq.~(\ref{xipower}). In the literature, the exponents associated with 
the specific heat, the magnetization, and the magnetic susceptibility are
denoted by $\alpha$, $\beta$, and $\gamma$, respectively. The exponent
of the magnetization at the critical temperature for a non-vanishing 
external field is denoted by $\delta$.  The exponent $\eta$ governs the 
behavior of the two-point function at the critical point. For the 
precise definition of these exponents and relations between them see for 
example section 1.3 of the review \cite{PeVi02}.
Second-order phase transitions are grouped into universality classes. For all
transitions within such a class, critical exponents assume identical values.
Also correction exponents such as $\theta=\omega \nu$ are universal.
Universality classes are characterized by the symmetry properties of the
order parameter at criticality, the range of the interaction and the spatial
dimension of the system. For reviews on critical phenomena see, for
example, \cite{WiKo,Fisher74,Fisher98,PeVi02}.

Note that in general the symmetry properties of the order parameter
can not be naively inferred from the microscopic properties of the system.
In particular, a symmetry might emerge that is not present in the
classical Hamiltonian. For example, in a binary mixture, the two components 
are not related by a $\mathbb{Z}_2$ symmetry. However, in case the mixing-demixing 
transition is of second order, it belongs to the Ising universality class, 
which is characterized by a $\mathbb{Z}_2$ symmetry of the order parameter.

In the present work we are aiming at a precise determination of the 
critical exponents of the three-dimensional Heisenberg universality class.
To this end, 
we study a generalized icosahedral model on the simple cubic lattice. 
The field variable takes the normalized vertices of the icosahedron as values.
In addition $(0,0,0)$ might be assumed. In the following we refer to field
variables assuming this value as voids.
The idea to use a discrete subset of the sphere as values of the field
variable is rather old \cite{Rapa85,MaOdPa}, however received little attention.
Note that the field variable is also referred to as spin.
The model has a parameter $D$ that controls the density of voids. 
For a certain range of this parameter, the model undergoes a
second-order phase transition. 
Our numerical data show that at the phase transition, the model
is in the domain of attraction of the $O(3)$-invariant fixed point.
Hence, the model shares the universality class of the three-dimensional 
$O(3)$-invariant Heisenberg model. On the critical line, similar to the 
Blume-Capel model, the amplitude of the leading correction depends on $D$.
We demonstrate that there is one value $D^*$ of $D$, where leading corrections
to scaling vanish. We refer to the model at $D \approx D^*$ as improved model.
Simulating an improved model allows us to get more accurate estimates of 
universal quantities at a given budget of CPU time. 
For a brief discussion and references on improved models in general 
see \cite{myClock}.

As discussed below in more detail, a perturbation with the
symmetry properties of the icosahedron is irrelevant at the $O(3)$-invariant
fixed point. In the Appendix \ref{omegaico} we determine the corresponding
renormalization group (RG) eigenvalue $y_{ico}=-2.19(2)$. 
Likely, analogous to the case of the clock model discussed in ref. 
\cite{myClock}, the perturbation is dangerously irrelevant. Meaning that
in the low-temperature phase, in the thermodynamic limit, the spontaneous
magnetization might only assume one of the $12$ directions that are
preferred by the Hamiltonian. Note, however, that this does not affect 
the finite size scaling study at the critical point that we perform here.

Our motivation to study this model is that simulations 
take less CPU time than for an $O(3)$-invariant model as, for example, the 
$\phi^4$ model simulated in refs. \cite{myO3O4,ourHeisen,HaVi11} and less 
memory is needed to store the field variables.
The idea of the present work is similar to that of ref. \cite{myClock}, 
where we studied the $(q+1)$-state clock model. In addition to the $q$ values
with unit length, the value $(0,0)$ might be assumed by the field variable.
In the case of the $(q+1)$-state clock model, an $O(2)$-invariant model
can be approached by taking the limit $q \rightarrow \infty$. In contrast,
here we are restricted to the Platonic solids.

The Heisenberg universality class describes the critical behavior of 
isotropic magnets, for instance the Curie transition in isotropic ferromagnets 
such as Ni and EuO, and of antiferromagnets such as RbMnF$_3$ at the N\'eel 
transition point. A summary of experimental results for critical exponents is given in 
the tables 24 and 25 of the review \cite{PeVi02}. An example for a more recent 
experimental study is ref. \cite{He18}.  In table 2 of \cite{He18} estimates for
the critical exponents $\beta$, $\gamma$, $\delta$ and $\alpha$ are presented 
for four different materials. To get an idea of the accuracy that is achieved
let us pick out two results for GdScGe: $\alpha = -0.134\pm 0.005$ for the 
exponent of the specific heat and $\delta=4.799\pm 0.006$ for the critical 
exponent of the magnetization on the critical isotherm.
Using scaling relations these exponents can be converted to $\nu=0.7113(17)$ 
and $\eta=0.0347(11)$, which are the exponents given in table 
\ref{methods} below.

The three-dimensional Heisenberg universality class has been studied 
by using various theoretical approaches.  Well established field theoretic
methods are the $\epsilon$-expansion and the perturbation theory 
in three dimensions fixed. 
In order to extract numerical estimates for critical exponents, various 
resummation schemes are discussed in the literature.
 As examples we give in table \ref{methods} the estimates 
obtained in ref. \cite{GuZi98}.  Recently there has been progress in the 
$\epsilon$-expansion and the six-loop coefficient has been computed for the 
$O(N)$-invariant $\phi^4$ theory \cite{KoPa17}.  In table \ref{methods}, 
we give the results of the resummation used in ref. \cite{KoPa17} based 
on the five- and six-loop $\epsilon$-expansion.  The five- and six-loop
estimates are consistent.  Note however that for the five-loop resummation, 
the estimate of the error differs at lot between ref. \cite{GuZi98} 
and ref. \cite{KoPa17}.
For a discussion of the resummation schemes used, we refer the reader to
refs. \cite{GuZi98,KoPa17}. The $\epsilon$-expansion has been 
extended to seven-loop \cite{Schnetz18}. However no numerical estimates 
for critical exponents have been computed so far.

Great progress has been achieved recently by using the so called
conformal bootstrap (CB) method. 
In particular in the case of the three-dimensional
Ising universality class, the accuracy that has been reached for critical
exponents clearly surpasses that of other theoretical methods. See ref.
\cite{Simmons-Duffin:2016wlq} and references therein. Very recently also
highly accurate estimates were obtained for the XY universality class
\cite{che19}, surpassing the accuracy of results obtained by lattice 
methods. Still for the Heisenberg universality class \cite{Kos:2016ysd}, 
the estimates are less precise than those obtained by other methods.

Considerable progress has also been achieved by using the functional 
renormalization group method. In ref. \cite{DePo20} the authors have 
computed the critical exponents $\nu$, $\eta$ and the correction exponent 
$\omega$ for various values of $N$.  In the tables IV, V, VI and VII
of \cite{DePo20} the authors summarize their results and compare 
them with estimates obtained by other methods for $N=1$, $2$, $3$, and $4$,
respectively. A good agreement with the results of the conformal
bootstrap is found. The same holds for the comparison with estimates
obtained by studying lattice models. In table \ref{methods} we report
the estimates obtained for $N=3$.

Finally we report results obtained for the $O(3)$-invariant 
$\phi^4$ model on the simple cubic lattice.
Note that there exists a value $\lambda^*$ of the coupling constant $\lambda$ 
of this model such that the leading correction to scaling vanishes.
In ref. \cite{myO3O4} a finite size scaling analysis of Monte Carlo (MC) data
was performed. 
In ref. \cite{ourHeisen} both Monte Carlo simulations
and the high temperature (HT) series expansion were used. In particular, the 
analysis of the HT series by using integral approximants \cite{Gu} is
biased by using the estimates of the inverse critical temperature and 
$\lambda^*=4.6(4)$ obtained by Monte Carlo simulations. 
In ref. \cite{HaVi11} we mainly focused on the RG-eigenvalues 
of anisotropic perturbations at the $O(N)$-invariant fixed point.
As a byproduct, we get the revised estimate $\lambda^*= 5.2(4)$.
The values quoted for refs. \cite{ourHeisen,HaVi11} are obtained
by inserting this value into eqs.~(13,14,19) of ref. \cite{ourHeisen}. 
Next we report the results of Monte Carlo simulations that we discuss
in appendix \ref{appendixA}. The estimate of the inverse critical 
temperature and $\lambda^*$ are used to bias the HT analysis of ref. 
\cite{ourHeisen}.
\begin{table}
\caption{\sl \label{methods}
We give a selection of theoretical results for the critical exponents $\nu$
and $\eta$ and the exponent $\omega$ of the leading correction to scaling
for the three-dimensional Heisenberg universality class obtained by various
methods. For a more comprehensive summary see for example table 23 of
ref. \cite{PeVi02}. For the definition of the acronyms and a discussion 
see the text.
}
\begin{center}
\begin{tabular}{ccclll}
\hline
  Ref. & method  & year &   \mc{1}{c}{$\nu$}  &  \mc{1}{c}{$\eta$} & \mc{1}{c}{$\omega$}  \\
\hline
\cite{GuZi98} & 3D-exp.         & 1998 & 0.7073(35) &0.0355(25) & 0.782(13) \\
\cite{GuZi98} &$\epsilon$-exp. 5l&1998 & 0.7045(55) &0.0375(45) & 0.794(18)\\
\cite{KoPa17} &$\epsilon$-exp. 5l&2017 & 0.7056(16) &0.0382(10) & 0.797(7)\\
\cite{KoPa17} &$\epsilon$-exp. 6l&2017 & 0.7059(20)& 0.0378(5) &  0.795(7)\\
\cite{Kos:2016ysd}& CB      &2016  &  0.7121(28) & 0.03856(124) & \mc{1}{c}{-} \\
\cite{DePo20}     & NRG     &2020  &  0.7114(9)  & 0.0376(13)   & 0.769(11) \\
\cite{myO3O4} &   MC          & 2001  &  0.710(2)   & 0.0380(10) &\mc{1}{c}{-} \\
\cite{ourHeisen} & MC+HT      & 2002  &  0.7112(5)  & 0.0375(5) &  \mc{1}{c}{-} \\
\cite{ourHeisen,HaVi11}& MC+HT& 2002  & 0.7117(5)  & 0.0378(5) &  \mc{1}{c}{-} \\
\cite{HaVi11}     & MC      &2011  &  0.7116(10) &0.0378(3)  & \mc{1}{c}{-}  \\
 \cite{ourHeisen}, present work & MC+HT, $\phi^4$ &2020 &0.7116(2) &0.0378(3)&   \mc{1}{c}{-} \\
 present work     & MC, $\phi^4$  &2020  & 0.71164(25)  & 0.03782(10)   & \mc{1}{c}{-}     \\
 present work     & MC, icosahedral   &2020  & 0.71164(10)   & 0.03784(5)  & 0.759(2) \\
\hline
\end{tabular}
\end{center}
\end{table}
Finally we report the results obtained from the finite size scaling study
of the generalized icosahedral model. By using a hybrid of local and
cluster algorithms we simulated lattices of a linear size up to $L=400$. 
It is virtually impossible to give a comprehensive summary of the vast
literature on the subject. For a more extensive summary see for example 
table 23 of ref. \cite{PeVi02}.

We notice that our results for $\nu$ and $\eta$ obtained for the 
generalized icosahedral model are fully consistent with
those that were obtained for the $\phi^4$ model on the simple cubic lattice.
Our results are also consistent with but more precise than those of refs. 
\cite{Kos:2016ysd,DePo20} obtained by using the conformal bootstrap method
and the functional renormalization group method, respectively.

Comparing with the results obtained from the resummation of the 
$\epsilon$-expansion we see clear differences. Our result for 
$\nu$ is larger than that obtained in ref. \cite{KoPa17} by about three 
times the error that is quoted. The result for the correction exponent 
$\omega$ obtained in ref. \cite{KoPa17} is roughly by five times the error
that is quoted larger than ours.

The outline of the manuscript is the following:
In section \ref{theModel} we define the  model and the observables that we
measured. Furthermore, we summarize theoretical results on subleading 
corrections to scaling.
In section \ref{theAlgo} we discuss the Monte Carlo algorithm used in
the simulations and outline our approach to the analysis of the data.
In section  \ref{DataAnalysis} we analyze the data and present
the results for the fixed point values of phenomenological couplings, inverse
critical temperatures, the correction exponent $\omega$, and the
critical exponents $\nu$ and $\eta$. In section \ref{summary} we conclude and 
give an outlook.  In Appendix \ref{appendixA} we discuss our results for the 
three-component $\phi^4$ model on the simple cubic lattice. Finally, in Appendix
\ref{omegaico} we determine the RG-exponent $y_{ico}$ related with the breaking 
of the $O(3)$ symmetry.  

\section{The model}
\label{theModel}
We consider a simple cubic lattice. A site is given by $x=(x_0,x_1,x_2)$,
where $x_i \in 0,1,2,...,L_i-1$. In our simulations $L_0=L_1=L_2=L$ throughout
and periodic boundary conditions are imposed.
The model is analogous to the $(q+1)$-state clock model discussed in 
ref. \cite{myClock}.  In the case of the $(q+1)$-state clock model the
spins $\vec{s}_x$ take either values on the unit circle or assume 
the value $(0, 0)$. 
Here the circle is replaced by the two-sphere. In particular, the spin 
$\vec{s}_x$ 
might take one of the thirteen values $\vec{v}_m$ tabulated below:
\begin{equation}
 \label{allowedspins}
(0,0,0) \;,\;\; z (0, \pm 1, \pm \phi) \;,\;\; z (\pm 1, \pm \phi, 0) \;,\;\;
z (\pm \phi, 0, \pm 1) \;,
\end{equation}
where $\phi= \frac{1}{2} (1 + \sqrt{5})$ is the golden ratio and 
$z=1/\sqrt{1+\phi^2}=1/\sqrt{2+\phi}$.
The twelve vectors with unit length are the normalized vertices of the 
icosahedron.  
See for example eq.~(A.20) of ref. \cite{Cara}, which is eq.~(40) of the
preprint version. 
An alternative choice is given in eq.~(A.9), 
corresponding to  eq.~(29) of the preprint version.
In our simulation program the field variables are stored by using the label
$m \in \{0,1,2,...,12\}$, where $\vec{v}_0=(0,0,0)$ and $m \in \{1,2,...,12\}$
are assigned to the vectors of unit length.

In the following we shall refer to the model as generalized icosahedral model.
The reduced Hamiltonian is given by
\begin{equation}
\label{ddXY}
 {\cal H} = -  \beta \sum_{\left<xy\right>}  \vec{s}_x \cdot
     \vec{s}_y -D  \sum_x \vec{s}_x^{\,2} - \vec{H} \sum_x \vec{s}_x \;,
\end{equation}
where $\left<xy\right>$ denotes a pair of nearest neighbor sites on 
the simple cubic lattice. We introduce the weight factor
\begin{equation}
\label{weight}
w(\vec{s}_x)  = \delta_{0,\vec{s}_x^{\,2}} +  \frac{1}{12} \delta_{1,\vec{s}_x^{\,2}}
\end{equation}
that gives equal weight to $(0,0,0)$ and the collection of the 12 values with 
$|\vec{s}_x| =1$. Now the partition function can be written as
\begin{equation}
 Z = \sum_{\{\vec{s}\} }  \prod_x w(\vec{s}_x) \; \exp(-{\cal H}) \;,
\end{equation}
where $\{\vec{s}\}$ denotes a configuration of the field.   

The reduced Hamiltonian~(\ref{ddXY}) and the weight~(\ref{weight}) are 
the same as for the $(q+1)$-clock model defined in section II
of ref. \cite{myClock}. The two models only differ in the set of allowed
values of the field variables.  Note that in the limit 
$D \rightarrow \infty$ the value $(0,0,0)$ is completely suppressed.
In the following we consider a vanishing external field $\vec{H} =(0,0,0)$
throughout.

\subsection{The quantities studied}
The most important 
quantities are dimensionless quantities $R_i$ that are also called 
phenomenological couplings.  In particular we study the ratio of 
partition functions $Z_a/Z_p$, where $a$ denotes a system with anti-periodic 
boundary conditions in one of the directions and periodic ones in the remaining 
two directions, while $p$ denotes a system with  periodic boundary conditions
in all directions.  Furthermore, we study the second moment correlation length
over the linear lattice size $\xi_{2nd}/L$, the Binder cumulant $U_4$ and 
its generalization $U_6$. The exponent of the correlation length is determined 
by studying the finite size scaling behavior of the slopes of 
dimensionless quantities. 
The critical exponent $\eta$ is obtained from  
the finite size scaling behavior of the magnetic susceptibility $\chi$. 
These quantities are defined for example in section II B of 
ref. \cite{myClock}. In our analysis, the 
observables are needed as a function of the inverse temperature $\beta$
for a neighborhood of the inverse critical temperature $\beta_c$. To this
end, we simulate at $\beta_s$, which is a preliminary estimate of $\beta_c$ 
and compute the coefficients of the Taylor expansion in $(\beta-\beta_s)$
up to third order.

\subsection{Subleading corrections to scaling}
\label{subleading}
Analyzing our data, we use prior information on 
subleading corrections to scaling. These corrections are due to 
$O(N)$-invariant perturbations of the fixed point and perturbations
that break the $O(N)$-invariance. Let us first discuss the former.
In section III A of ref. \cite{myClock} we conclude,
based on the literature, that there should be only a small
dependence of the irrelevant RG-eigenvalues on $N$.  Therefore 
the discussion of section III A of ref. \cite{myClock} should 
apply to the present case $N=3$ at least on a qualitative level. 
In particular, we regard the subleading correction 
exponent $\omega_2=1.78(11)$ that we assumed in refs. 
\cite{ourHeisen,HaVi11} as an artifact of the scaling field 
method \cite{NewmanRiedel}.
Instead, the most important subleading correction should be 
due to the breaking of the rotational symmetry by the simple
cubic lattice. Following ref. \cite{ROT98}, the associated
correction exponent is $\omega_{NR} \approx 2.02$.

Now let us turn to the corrections caused by the breaking 
of the $O(3)$-invariance. A good starting point of the discussion 
is provided by ref. \cite{Cara}.  In section 2, polynomials are
constructed that are invariant under the action of the discrete
symmetry groups related with the Platonic solids
 and belong to an irreducible representation of 
the $O(3)$ group. Hence, they have a well defined $O(3)$ spin $n$. 
In eq.~(3) of ref. \cite{Cara}, polynomials associated with the
tetrahedron, the cube, and the icosahedron are given. 
These are associated with the spin $n=3$, $4$, and $6$.  
Note that the tetrahedron is self-dual, the octahedron is dual
to the cube and the dodecahedron is dual to the icosahedron.
There are no further Platonic solids in three dimensions.
Note that dual Platonic solids share the symmetry properties.
Hence, using the  dodecahedron instead of the icosahedron as 
approximation of the sphere should result in the same irrelevant
RG-exponent.

In the case of a two-dimensional system, as discussed 
in ref. \cite{Cara}, these perturbations of the $O(3)$-symmetry
are relevant.
In particular, the icosahedral model undergoes a phase 
transition at a finite temperature, while the 
$O(3)$-symmetric model is asymptotically free
and hence no phase transition occurs at a finite temperature.

In ref. \cite{HaVi11} we determined the RG-exponents 
$y_n=1.7906(3)$, $0.9616(10)$, and $0.013(4)$ for $N=3$ and three
spatial dimensions for spin $n=2$, $3$, and $4$, respectively.
Hence, for example a cubical model could not be used to 
study the properties of the $O(3)$ invariant fixed point, since the 
perturbation is relevant.
We could not find a result for $N=3$ and $n=6$ in the literature.
However it is interesting to note that the estimates of $y_n$ for
$n=2$, $3$, and $4$ for
$N=3$ are well approximated by the average of the corresponding 
values for $N=2$ and $4$. In refs. \cite{Debasish2,Debasish4} 
the estimates $y_6=-2.509(7)$ and $-2.069(7)$ are given for $N=2$ and $4$, 
respectively. Therefore we would expect $y_6 \approx -2.29$ for $N=3$. 
In appendix \ref{omegaico} we find $y_{6}= -\omega_{ico} =  - 2.19(2)$. 
For a discussion of Platonic solids related with stable fixed points in 
three dimensions, see ref. \cite{Gori}. 

There are also corrections that are not related to irrelevant scaling 
field, such as the analytic background of the magnetic susceptibility.
Effectively, it behaves as a correction with the exponent $2-\eta$. 
For a more comprehensive discussion of subleading corrections see section III
of \cite{myClock}.

\section{Simulation algorithm}
\label{theAlgo}
The algorithm used is very similar to the one discussed 
in section IV of ref. \cite{myClock}. 
We simulated the model by using a hybrid of local updates and
cluster updates \cite{SW}. In the case of the cluster algorithm, we 
have implemented the single cluster algorithm \cite{Wolff} and the wall 
cluster algorithm \cite{wall}. 

\subsection{Local Metropolis updates}
\label{localAlg}
In order to speed up the local updates, in ref. \cite{myClock} we 
tabulate the contribution to the Boltzmann factor
by pairs
\begin{equation}
 B(m,n) = \exp(\beta \; \vec{s}(m) \cdot \vec{s}(n) )
\end{equation}
and its inverse $B^{-1}(m,n)$, where $m$ and $n$ are the labels of the 
values of the spins. In order to adapt the implementation of the 
local updates of ref. \cite{myClock} to the present case, we just had to 
plug in the scalar products $\vec{s}(m) \cdot \vec{s}(n)$ for the 
vectors given in eq.~(\ref{allowedspins}).

Similar to ref. \cite{myClock} we have used two versions of the 
Metropolis update that differ in the choice of the proposal. In the first
version we always propose $\vec{s}_x\,'=(0,0,0)$ if $|\vec{s}_x|=1$ and,
with equal probability, one of the 12 values with unit length if
$\vec{s}_x=(0,0,0)$.  

In the second version, the proposal does not depend on $\vec{s}_x$. 
With probability $1/2$ we propose $\vec{s}_x\,'=(0,0,0)$ and with 
probability $1/24$ one of the 12 values with unit length.  
The second choice is used in addition to the first one, since we were not
able to prove ergodicity for the first one.

\subsection{The cluster algorithms}
\label{clusteralg}
Using the cluster algorithm, a spin is potentially
changed  by a reflection at one of the 15 symmetry
planes of the icosahedron. The reflection can be written as
\begin{equation}
\label{reflection}
 \vec{s}\,' = \vec{s} - 2 ( \vec{r} \cdot \vec{s} \, ) \vec{r} \;,
\end{equation}
where $\vec{r}$ is a unit vector perpendicular to the symmetry plane. 
Being too lazy to 
search the literature, we computed the 
possible values of $\vec{r}$ by using a simple Python program. First we
define for all pairs of vertices $\vec{v}_i$ of the icosahedron a candidate
\begin{equation}
\vec{c}_{ij} = \frac{\vec{v}_i + \vec{v}_j}{|\vec{v}_i + \vec{v}_j|} \;.
\end{equation} 
Then we checked that the candidate is indeed a reflection. Finally
we search for multiple identifications of the same reflection. The 
remaining results for $\vec{r}$ can be grouped in 5 triples of vectors that 
are mutually orthogonal:
\begin{center} $ \begin{array}{ccc}
   (1,0,0)  & \;\;  (0,1,0) \;\; &  (0,0,1) \\
   (-a,1/2,b) & \;\;  (1/2,b,a) \;\;   &  (b,a,-1/2) \\
   (a,1/2,b)  & \;\;   (-1/2,b,a)  \;\; &   (b,-a,1/2) \\
   (1/2,b,-a) & \;\;  (b,a,1/2) \;\; & (a,-1/2,b) \\
   (-b,a,1/2) & \;\;  (1/2,-b,a) \;\;  & (a,1/2,-b) \;,
\end{array} $ \end{center}
 where $a=\phi/2$, $b=\phi/2-1/2$ and
 $\phi= \frac{1}{2} (1 + \sqrt{5})$ is the golden ratio.

As usual, the cluster algorithm is characterized by the delete probability
of a pair of nearest neighbor sites \cite{Wolff} 
\begin{equation}
\label{pdel}
 p_d(x,y) = \mbox{min}[1,
\exp (-2 \beta [\vec{r}  \cdot \vec{s}_x] [\vec{r}  \cdot \vec{s}_y] )]\;.
\end{equation}
Below we shall refer to a pair of nearest neighbor sites as link. A link 
$<xy>$ is deleted with probability $p_d(x,y)$. Otherwise it is frozen.

In the program, we computed all possible values of $p_d(x,y)$ before 
the simulation is started, and store the results in a $15 \times 13 \times 13$
array of double precision floating point values. 

Different cluster algorithms are characterized by the way clusters are 
selected. In the Swendsen-Wang algorithm \cite{SW}, the whole lattice is 
decomposed 
into clusters of sites that are connected by frozen links. In the Swendsen-Wang
algorithm, a cluster is flipped with probability $1/2$. Flipping means that
for all sites within a cluster, the reflection, eq.~(\ref{reflection}), is 
performed. In the case of the single cluster algorithm \cite{Wolff}, 
one site of the lattice is 
randomly selected. Then only the cluster that contains this site is 
constructed. This cluster is flipped with probability $1$. 
In the wall cluster algorithm \cite{wall}, instead of a single site a plane 
perpendicular to one of the lattice axis is chosen. The position on this axis 
is randomly chosen. 
Then all clusters that contain sites within this plane are 
constructed and flipped with probability one.  
The measurement of $Z_a/Z_p$ is discussed in the Appendix A 2 
of ref. \cite{XYold}.

\subsection{The update cycle}
The update steps discussed above are compounded into a complete update
cycle. Below we give a piece of pseudo \verb+C+-code that represents the cycle 
that is used in our simulations:

\begin{verbatim}

Metropolis_2();
for(k=0;k<3;k++)
  {
  Metropolis_1();
  ir=5*rand();
  wall_cluster((k+1)%3,triples[ir][0]);
  wall_cluster((k+1)%3,triples[ir][1]);
  wall_cluster((k+1)%3,triples[ir][2]);
  Metropolis_1();
  for(j=0;j<L;j++) single_cluster();
  Metropolis_1();
  ir=5*rand();
  wall_cluster_measure(k%3,triples[ir][0]);
  wall_cluster_measure(k%3,triples[ir][1]);
  wall_cluster_measure(k%3,triples[ir][2]);
  measurements();
  }
\end{verbatim}

Here \verb+Metropolis_1()+ and \verb+Metropolis_2()+ are sweeps, using
the first and second type of the Metropolis update discussed
in section \ref{localAlg}.
The single cluster update is given by  \verb+single_cluster()+. For each
call, the reflection $\vec{r}$ and the site, where the cluster is started
are randomly selected with a uniform distribution.
\verb+wall_cluster(k%3,triples[ir][i])+ is a wall cluster update. 
The first argument selects the spatial direction.
The array \verb+triples[ir][i]+ 
determines which reflection $\vec{r}$ is chosen for the cluster update.
The first index \verb+ir+ selects the set of mutually orthogonal 
$\vec{r}$ that is taken. Then within such a set we run through all three 
$\vec{r}$.  The wall cluster update is either called just for updating 
the configuration or, in the case of \verb+wall_cluster_measure+ to 
perform a measurement of the ratio of partition functions $Z_a/Z_p$ 
in addition.

Most of the simulations were performed by using the update cycle discussed
above. Below we shall refer to this cycle as cycle A. 
At a certain stage of the simulation, we realized that for some of the 
quantities it is more efficient to measure more frequently. Therefore we 
skipped the wall cluster updates without measurement and reduced the 
number of single cluster updates from $L$ to $L/2$.  Furthermore one
of the \verb+Metropolis_1()+ sweeps is skipped. Below we shall refer to this
cycle as cycle B.

We implemented the code in standard C and used the SIMD-oriented Fast Mersenne
Twister algorithm \cite{twister} as random number generator.

Since the program is essentially the same as the one used to simulate 
the $(q+1)$-state clock model, the CPU-times needed for the update of a single
site, are identical to those quoted in section IV C of ref. \cite{myClock}:
Our Metropolis update type one requires $1.2 \times 10^{-8}$ s per site. In the case
of the single cluster update about  $3.8 \times 10^{-8}$ s per site are needed.
These timings refer to running the program on a single core of an Intel(R) 
Xeon(R) CPU E3-1225 v3.
Compared with the simulation of the $O(3)$-symmetric $\phi^4$ model on the  
simple cubic lattice discussed below in appendix \ref{appendixA}, we roughly 
gain a factor of three.

\subsection{General remarks on the analysis of the data}
The quantities that we study follow a power law that is subject to corrections
\begin{equation}
\label{generalform}
 A(L)  = a L^{u} \;(1 + \sum_i c_i L^{-\epsilon_i}) \; ,
\end{equation}
where $L$ is the linear size of the lattice. By using Monte Carlo
simulations, we obtain estimates of $A(L)$ that have statistical errors.
Mostly we intend to determine the exponent $u$, which is either the RG-exponent
of the thermal scaling field $y_t=1/\nu$ or $2-\eta$ here. 
The amplitudes $a$ and $c_i$ 
are in general unknown. In the case of the correction exponents we have some
prior knowledge. This is gained by theoretical considerations or the 
analysis of other numerical data, as discussed in section \ref{subleading}
above.
We denote the correction exponents in eq.~(\ref{generalform}) by $\epsilon_i$,
since not all are related to a single irrelevant scaling field.
The correction exponents of irrelevant scaling fields are given by irrelevant  
RG-exponents $\omega_i =-y_i$.
Performing least-square fits, we need ans\"atze that contain only a few 
free parameters. Hence, the series of corrections in eq.~(\ref{generalform}) 
has to be truncated.  In our case there is the leading correction with 
the exponent $\omega = 0.759(2)$, see eq.~(\ref{omegafinal}) below. 
Extracting the critical exponents $\nu$ and $\eta$, we consider $D \approx D^*$
and on top of that improved observables that are constructed such that 
the leading correction is suppressed. Therefore it is safe to ignore the
leading correction.  As discussed in section \ref{subleading} there are a 
number of different corrections with $\epsilon_i \approx 2$. These are 
the analytic background of the magnetic susceptibility that effectively 
corresponds to $\epsilon_1 = 2 -\eta$, the violation of the rotational symmetry
by the simple cubic lattice $\omega_{NR} \approx 2.02$, and 
$\omega_{ico}=2.19(2)$ related to the breaking of the $O(3)$ symmetry.
In the case of the slopes that are used to determine $\nu$ there is also 
$\omega+1/\nu \approx 2.164$. In principle there is an infinite series of
corrections with increasing correction exponents. Since we can deal only
with a few free parameters in fits, the sequence has to be truncated at some
stage. 
Even the different corrections with an exponent $\epsilon_i \approx 2$ have
to be represented by a single or by two effective correction terms.
Hence in general the ansatz will never perfectly represent the
data. Therefore in addition to the statistical error there is a systematic
one that is caused by this imperfection.  With increasing linear lattice
size $L$, the magnitude of corrections decreases. If one would consider 
the linear lattice sizes $L_{min} \le L  \le c L_{min}$, where $c>1$, then
the estimate of the exponent $u$ would converge with increasing $L_{min}$, 
up to the statistical error, to the true answer. Of course, the CPU time 
that is available sets an upper limit to $c L_{min}$. 
Since we would like to squeeze out most from the data we proceed
in a different way, similar to most analyses in the literature,
all data with $L \ge L_{min}$ are taken into account.  The quality of the 
fit is measured as usual by 
\begin{equation}
\label{chiunco}
 \chi^2 = \sum_j [(f(x_j,\{p\})-y_j)/\sigma_j]^2 \;,
\end{equation}
where $f$ is the ansatz and $\{p\}$ the parameters of the ansatz. In our 
case, $x_j$ are the linear lattice sizes, $y_j$ the numerical estimates
of the observable and $\sigma_j$ its statistical error.
In some of the fits below we consider several observables jointly.  In this
case
\begin{equation}
\label{chico}
\chi^2 = r C^{-1} r^T \;,
\end{equation}
where $C$ is the covariance matrix and $r_j = y_j -f(x_j,\{p\})$. Note that
now $x_j$ refers to the linear lattice size and the type of the observable. 
We also perform joint fits for several values of $D$. Then $x_j$ also refers
to $D$. A fit usually is regarded as acceptable if $\chi^2/$d.o.f.$\approx 1$,
where d.o.f. is the number of degrees of freedom.
Furthermore we consider the goodness-of-fit
$Q=\Gamma_{inc}^{up}(d.o.f./2,\chi^2/2)$, where 
$\Gamma_{inc}^{up}$ is the regularized upper incomplete gamma-function.
For a Gaussian distribution of the numerical estimates $y_j$, $Q$ gives 
the probability that, assuming that the ansatz is correct, $\chi^2$ is equal
to or larger than the value that we find for our data.

Here we are dealing with ans\"atze that are only correct up to corrections
that decay with a power of the linear lattice size $L$. 
As a result, taking into account the smallest
$L$ that we have simulated, $\chi^2/$d.o.f. is large and $Q$ very small. 
Increasing
$L_{min}$, typically $\chi^2/$d.o.f. decreases and $Q$ increases. In 
all cases discussed below, eventually acceptable values of $\chi^2/$d.o.f. 
and $Q$ are reached. In our plots below we give only estimates that correspond 
to $Q>0.01$. Typically $Q$ rapidly increases going to slightly larger $L_{min}$.
For most of the estimates shown $Q>0.1$. 
A large value of $\chi^2/$d.o.f. or a small value of $Q$ certainly 
indicates that the ansatz that is used is not sufficient to describe the 
data. Unfortunately, however an acceptable value of  $\chi^2/$d.o.f. or $Q$
says little about the systematic error on the parameters such as the exponent
$u$.  In particular the systematic error can be considerably larger than the 
statistical one that is provided by the fit. This can be seen explicitly 
for example in our data for the slopes of different phenomenological 
couplings.  While the correction exponents are the same for different 
quantities, very likely the corresponding amplitudes are not. Hence the 
systematic effect on, for example, the result for the exponent $y_t$ is 
likely different for different quantities. And in fact we see differences 
in $y_t$ obtained from different quantities that are clearly larger 
than the statistical error, despite the fact that $Q$ is acceptable. 
This effect can also be easily seen by generating synthetic data according
to a function $g$ with given values of the parameters and then fitting by using
the ansatz $f$, where $f$ is obtained from $g$ by skipping correction terms. 

In order to get some handle on the systematic error we compare results
obtained by the same ansatz but different quantities or by different ans\"atze,
containing a different number of correction terms for the same quantity.
The final analysis is performed graphically. We plot the estimate of, for
example, $y_t$ as a function of $L_{min}$. The final result and its error
is then chosen such that for all quantities or all ans\"atze 
considered the estimate obtained by fitting is, including the respective 
statistical error, within the interval given by the final estimate plus or
minus its error. This procedure is not fully automatized and subject to 
some judgment. 
 
The least square fits were performed by using the function {\sl curve\_fit()}
contained in the SciPy  library \cite{pythonSciPy}. The function
{\sl curve\_fit()} acts as a wrapper to functions contained in the
MINPACK library \cite{MINPACK}. We checked the outcome of the 
fit by varying the initial values of the parameters. Furthermore, we performed 
fits both by using the Levenberg-Marquardt algorithm and the trust region 
reflective algorithm. In particular in the case of fits with many free
parameters, the trust region reflective algorithm turns out to be more reliable 
than the Levenberg-Marquardt algorithm. 
Plots were generated by using the Matplotlib library \cite{plotting}.

\section{The simulations}
\label{DataAnalysis}
Our simulations were performed on various PCs and servers. The CPU
times quoted below refer to a single core of an Intel(R) Xeon(R) 
CPU E3-1225 v3 running at 3.20 GHz, which is the CPU of our PC at home.
For example for an AMD EPYC$^{TM}$ 7351P CPU we find very similar times,
running the program on a single core.

First we performed a number of preliminary simulations to map out the
phase diagram of the model. There is a line of
second-order phase transitions that starts at $D=\infty$ extending to 
$D_{tri} \approx -0.5$.  For smaller values of $D$, the transition is
of first order. Our preliminary estimate for the improved model is
$D^* \approx 2$. We also obtained preliminary estimates of the inverse
critical temperature $\beta_c(D)$ for various values of $D$. 
Based on these preliminary results we  arranged our main simulations.

For $D=2.05$ and $2.1$ we simulated the linear lattice sizes 
$L=4$, $5$, ..., $14$, $16$, ..., $24$, $28$, ..., $48$, $56$, ..., $80$, 
$90$, $100$, $140$, $200$, and $400$.
In the case of $D=2.0$ we simulated the same lattice sizes up to $L=64$.
Larger lattice sizes are only $L=80$ and $200$. For example for $D=2.1$ we
performed about $3  \times 10^9$  measurements up to $L=32$. Then the 
statistics is slowly decreasing to $6.6 \times 10^8$ measurements 
for $L=100$. We performed $3.5  \times 10^8$, $1.45 \times 10^8$, and 
$1.8 \times 10^7$ measurements for $L=140$, $200$, and $400$.  Most of the
simulations were performed by using cycle A. For $L=90$ and $400$, cycle B
was used.

The simulations for $D=2.0$, $2.05$ and $2.1$ took in total 
60 years of CPU time.
These simulations were performed to accurately determine the fixed 
point values $R_i^*$ of dimensionless quantities and $D^*$. 
The critical exponents $\nu$ and $\eta$  are determined by using data
generated for $D=2.05$ and $2.1$.

In addition we simulated at $D=\infty$,
$1.4$, $1.0$, $0.5$, $0.0$, and $-0.3$ using lattice sizes up to $L=90$. 
This set of simulations mainly serves to determine the exponent of leading 
corrections to scaling $\omega$. Furthermore improved observables are 
constructed based on these data.
Also for these simulations, we spent in total 60 years of CPU time.

\subsection{Fixed point values of the RG-invariant quantities and critical 
temperatures}
\label{fixedpoint}

In this section, we determine the critical temperature
for $D=2.0$, $2.05$, and $2.1$, $D^*$ and the fixed
point values of phenomenological couplings. First we analyze the 
phenomenological couplings one by one, similar to the analysis performed
in section V A of ref. \cite{myClock}.
We use the ans\"atze  
\begin{eqnarray}
 R_i(L,D,\beta_c(D)) &=& R_i^* \;, \label{RR0} \\
 R_i(L,D,\beta_c(D)) &=& R_i^* + b_i(D) L^{-\epsilon_1} \;, \label{RR1} \\
 R_i(L,D,\beta_c(D)) &=& R_i^* + b_i(D) L^{-\epsilon_1} 
                             + c_i(D) L^{-\epsilon_2} \;, \label{RR2} \\
 R_i(L,D,\beta_c(D)) &=& R_i^* + b_i(D) L^{-\epsilon_1} 
                             + c_i(D) L^{-\epsilon_2}
                             + d_i(D) L^{-\epsilon_3} \;,
\label{RR3}
\end{eqnarray}
where we approximate 
\begin{equation}
\label{bs}
 b_i(D) = b_{s,i} \; (D-D^*) 
\end{equation}
linearly and $c_i(D)$ and $d_i(D)$ being constant for $D=2.0$, $2.05$ and $2.1$ 
that we consider here. Note that writing the leading correction amplitude
$b_i(D)$ this way, $D^*$ is an immediate parameter of the fit. Also the slopes
$b_{s,i}$ contain important information as we shall see below, 
eq.~(\ref{imporvedR}). We take $\epsilon_1=0.76$, $\epsilon_2=2$, and either
$\epsilon_3=2.2$ or $\epsilon_3=4$. Note that $\epsilon_1$ is close to the 
estimate $\omega=0.759(2)$ obtained below. The analytic background of
the magnetic susceptibility and the violation of the rotational invariance 
are effectively taken into account by the term $c L^{-\epsilon_2}$.  The 
choice $\epsilon_3=2.2$ is motivated by a preliminary estimate of 
$\omega_{ico}$. We checked that taking for example $\epsilon_3=2.17$ instead, 
changes the estimates of the critical temperature and $R_i^*$ by little.
Adding a term $c L^{-4}$ is mainly driven by the observation that this way
$\chi^2/$d.o.f. $\approx 1$ are obtained down to $L_{min}=5$. This observation
suggests that there is a correction with an RG-exponent $y\approx -4$ that 
has a quite large amplitude. Also analyzing different quantities we find
that adding a term $c L^{-4}$ results in acceptable fits 
down to $L_{min}=5$. 

Our final results are mainly based on fits with two correction terms, 
eq.~(\ref{RR2}). 
Other fits serve to estimate systematic errors.
Our results are summarized in table \ref{betac1}.

\begin{table}
\caption{\sl \label{betac1}
In the first column the phenomenological coupling is specified.
In the second column we give the corresponding
estimates of the fixed point values $R^*$ obtained by separate fits
for each  phenomenological coupling. In the third column we give
the estimates of the fixed point values $R^*$ obtained by joint fits 
of all four  phenomenological couplings.
In the fourth column we give the estimates of $D^*$, where leading
corrections to scaling vanish.  In the following columns, the estimates
of the inverse critical temperature $\beta_c$ for $D=2.0$, $2.05$, and
$2.1$ are given.  In rows one to four we give the estimates obtained 
by fitting the phenomenological coupling separately, while 
in the last row we give  estimates obtained from joint fits.
}
\begin{center}
\begin{tabular}{clllllll}
\hline
\multicolumn{1}{c}{$R$}&\multicolumn{1}{c}{$R^*_{sep}$} &\multicolumn{1}{c}{$R^*_{joint}$}
& \multicolumn{1}{c}{$D^*$} &
\multicolumn{1}{c}{$\beta_c(2.0)$} & \multicolumn{1}{c}{$\beta_c(2.05)$} &
\multicolumn{1}{c}{$\beta_c(2.1)$} \\
\hline
$Z_a/Z_p$ &0.19479(6)&0.19477(2) & 2.1(1)&0.74542805(10)  & 0.74296024(7) &0.74060257(7)  \\
$\xi_{2nd}/L$&0.564005(30) &0.56404(2) &2.14(5)&0.74542795(8)&0.74296021(6) & 0.74060251(6) \\
$U_4$      &1.13933(4) & 1.13929(2) & 2.06(3) & 0.74542800(9) & 0.74296018(8) &0.74060255(8) \\ 
$U_6$      &1.41985(15) & 1.41974(5) & 2.06(3) &0.74542800(9)  & 0.74296018(8)& 0.74060255(8)  \\
\hline
joint         &       &   & 2.08(2)  &0.74542801(5)  & 0.74296024(5)   & 0.74060256(5) \\
\hline
\end{tabular}
\end{center}
\end{table}

In contrast to previous work \cite{myClock}, 
we made an attempt to jointly fit all four phenomenological
couplings $R$ that we consider. To this end we computed the covariances of the
different $R$.  Since only quantities with the same $D$ and $L$ are correlated,
the covariance matrix is sparse. Only four by four blocks are non-vanishing.
For example for $L_{min}=8$, there are $22 + 27 + 27$ different $(D,L)$ pairs. 
Hence the covariance matrix is a 
$[ 4 \; (22 + 27 + 27) ] \times [ 4 \; (22 + 27 + 27) ]$ matrix. 
We passed the full $[ 4 \; (22 + 27 + 27) ] \times [ 4 \; (22 + 27 + 27) ]$ 
covariance matrix to  \verb+optimize.curve_fit+, since we found no
simple way to indicate that the matrix is sparse. Since the optimization 
typically  took a few seconds, we made no effort to improve on this. 

It turns out that in these joint fits, we can include more 
correction terms. As above, we fixed $\epsilon_1=0.76$ corresponding to
the exponent of the leading correction $\omega$.
We consider the sequence  $\epsilon_i = 2 -\eta$, $2.02$, and $2.19$ of 
subleading corrections exponents.

We find that $Q>0.1$ 
for $L_{min} \ge 22$, $11$, and $9$, taking into account 1, 2 or 3
subleading correction terms, respectively. Here, adding a correction 
$\propto L^{-4}$ does not improve the fits much.

\begin{figure}
\begin{center}
\includegraphics[width=14.5cm]{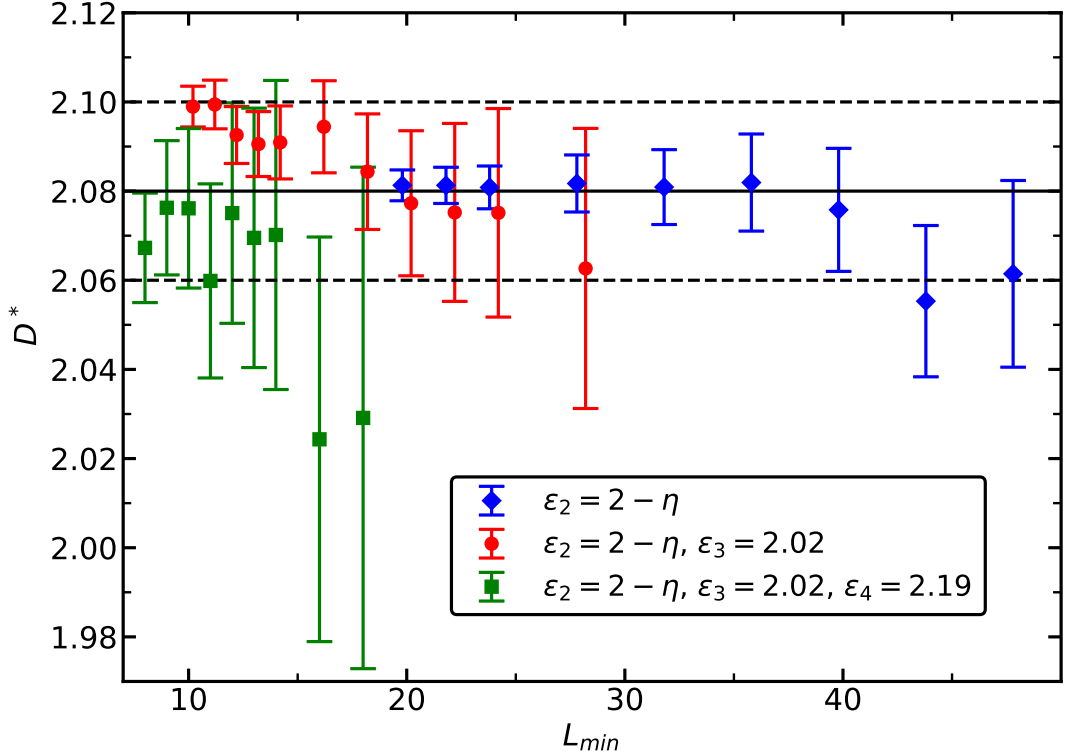}
\caption{\label{Dsjoint}
We plot the estimate of $D^*$ obtained by using joint fits of phenomenological
couplings as function of the minimal lattice size $L_{min}$ that is 
taken into account. Data for $D=2.0$, $2.05$ and $2.1$ are used in the fits.
Only results for $Q > 0.01$ are given. To make the figure more
readable, we have shifted the values of $L_{min}$ slightly.
The solid line indicates the preliminary estimate based on this set of fits.
The dashed lines give the error estimate.
The exponents $\epsilon_i$ given in the legend refer to the correction terms
that are included in the ans\"atze.  
}
\end{center}
\end{figure}

In Fig. \ref{Dsjoint} we give the results
obtained for $D^*$ by using these fits as a function of the minimal 
lattice size $L_{min}$. In the plot, only results that correspond to $Q > 0.01$
are given.
The final estimate of $D^*$ and its error bar are chosen such that the 
estimates of $D^*$ obtained by the individual fits, including
their respective error bars, are contained in the interval that is given
by the final estimate plus or minus its error for some range of $L_{min}$.
For example, the results for the 
ansatz only containing the subleading correction term $\propto L^{-2+\eta}$ 
are within this interval up to $L_{min}=40$. 

In a similar fashion we determine the final estimates of the fixed 
point values of the phenomenological couplings and the inverse critical 
temperatures. These are summarized in table \ref{betac1}.

\subsubsection{Eliminating leading corrections to scaling in dimensionless
 quantities}
We construct linear combinations of two phenomenological couplings 
\begin{equation}
\label{imporvedR}
 R_{imp,i,j} = R_i + p_{i,j} R_j  \;,
\end{equation}
such that leading corrections to scaling are eliminated. To this end, we 
make use of the parameter $b_{s,i}$, eq.~(\ref{bs}), which we determined 
in the analysis of the phenomenological couplings discussed above. One gets
\begin{equation}
\label{imporvedRb}
 p_{i,j} = - \frac{b_{s,i}}{b_{s,j}} \;. 
\end{equation}
Note that these results hold for any model in the three-dimensional 
Heisenberg universality class. Here we consider the combination of either 
$Z_a/Z_p$ or $\xi_{2nd}/L$ with $U_4$. Our numerical estimates are 
summarized in table \ref{pR}.

\begin{table}
\caption{\sl \label{pR}
Estimates of the coefficient $p_{i,j}$, eq.~(\ref{imporvedR}), needed to 
construct improved phenomenological couplings.
}
\begin{center}
\begin{tabular}{ccl}
\hline
   $R_i$  &  $R_j$ &  \multicolumn{1}{c}{$p_{i,j}$} \\
\hline
$Z_a/Z_p$    & $U_4$  &  0.575(25)   \\  
$\xi_{2nd}/L$ & $U_4$  & -0.750(25)  \\
\hline
\end{tabular}
\end{center}
\end{table}
Jointly fitting the data for $D=2.0$, $2.05$ and $2.1$, assuming that 
leading corrections to scaling vanish, we find 
$(Z_a/Z_p+0.575 \; U_4)^* = 0.84987(3)$
and $(\xi_{2nd}/L-0.75 \; U_4)^* = -0.290437(10)$.  Note that these results 
are consistent with those obtained by naively combining the estimates of $R^*$
given in table \ref{betac1}.
Since the leading correction to scaling is eliminated up to the numerical 
uncertainty of the coefficients $p_{ij}$, these linear combinations are well
suited to  determine the inverse critical temperature $\beta_c$ of models that  
are not improved. Furthermore, in the slope of these combinations the effective 
correction $\propto L^{-y_t  - \omega}$ is eliminated. In the analysis of
the generalized icosahedral model we shall not make use of this fact, since 
$\omega_{ico}$ assumes a value that is similar to $y_t+\omega$. 

In table \ref{betacviel} we give estimates of $\beta_c$ obtained 
by analyzing $Z_a/Z_p+0.575 \; U_4$. We use the estimate of 
$(Z_a/Z_p+0.575 \; U_4)^*$ given above as input. The error quoted also 
takes into account the uncertainty of $(Z_a/Z_p+0.575 \; U_4)^*$. 

\begin{table}
\caption{\sl \label{betacviel}
Estimates of $\beta_c$ for the values of $D$ different from 
$D=2.0$, $2.05$, $2.1$. 
}
\begin{center}
\begin{tabular}{cc}
\hline
  $D$     &    $\beta_c$ \\
\hline
$\infty$& 0.6925051(2) \\
1.4  &    0.7854535(2) \\
1.0  &    0.8260052(2)  \\
0.5  &    0.8979286(2)  \\
0.0  &    0.9986988(2)  \\
-0.3  &   1.0742253(4)  \\
\hline
\end{tabular}
\end{center}
\end{table}

Note that the value of $\beta_c$ for $D=\infty$ is slightly smaller than
$\beta_c=0.693002(2)$ for the $O(3)$-invariant Heisenberg model 
on the simple cubic lattice \cite{Deng05}. 

\subsection{Leading corrections to scaling}
\label{corrections}
In this section we focus on leading corrections to scaling. To this end
we consider the cumulants $U_4$ and $U_6$ at 
$Z_a/Z_p=0.19477$ or $\xi_{2nd}/L=0.56404$, which are our estimates
of the fixed point values of these quantities. This means that
$U_4$ and $U_6$ are taken at $\beta_f$, where $\beta_f$ is chosen such that
either $Z_a/Z_p=0.19477$ or $\xi_{2nd}/L=0.56404$. In the following
we denote a cumulant at a fixed value of  $Z_a/Z_p$ or $\xi_{2nd}/L$
by $\bar{U}$. 

\begin{figure}
\begin{center}
\includegraphics[width=14.5cm]{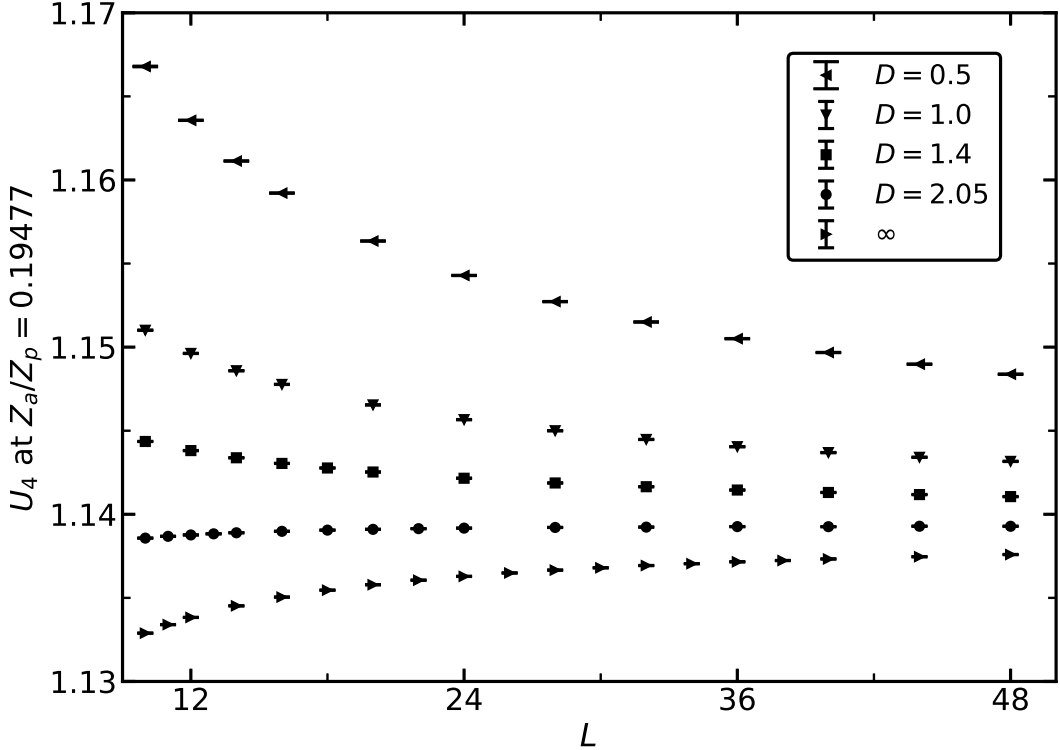}
\caption{\label{U4Z}
We plot $U_4$ at $Z_a/Z_p=0.19477$ for $D=0.5$, $1.0$, $1.4$, $D=2.05$ 
and $\infty$ as a function of the linear lattice size $L$.
}
\end{center}
\end{figure}

To get a first impression, we plot in Fig. \ref{U4Z} the Binder cumulant
$U_4$ at $Z_a/Z_p=0.19477$ for $D=0.5$, $1.0$, $1.4$, $2.05$ and $\infty$. 
To keep the figure readable, we do not plot the data for $D=2.0$ and 
$2.1$, which are similar to those of $D=2.05$. For $D=1.4$ the correction
amplitude has roughly the same modulus as for $D=\infty$, but opposite sign. 
The amplitude of leading corrections increases with decreasing $D$.  
The analysis performed below shows that the amplitudes of leading corrections 
for $D=0.0$ and $-0.3$ are about $2.7$ and $7$ times as large as for $D=0.5$,
respectively. The results obtained for $\xi_{2nd}/L=0.56404$ are similar.
The results obtained for $U_6$ are qualitatively the same as for $U_4$.

We performed joint fits for several sets of $D$. Either all values 
of $D$ are taken into account, or subsets of them.
These subsets are obtained by skipping values of $D$ starting from the 
smallest one. The minimal set that we consider consists of 
$D=1.4$, $2.0$, $2.05$, $2.1$, and $\infty$.
Similar to ref. \cite{myClock} we analyzed our data by using the ansatz
\begin{equation}
\label{manyDfit}
\bar{U} = \bar{U}^* + \sum_{i=1}^{i_{max}} c_i [b(D) L^{-\omega}]^i  + d L^{-\epsilon}
\end{equation}
for various values of $i_{max}$. In order to avoid ambiguity, we set $c_1=1$.
The free parameters of the fit are $\bar{U}^*$, $c_2$, $c_3$, ..., $b(D)$ for
each value of $D$. In our fits,  the parameter $d$ is 
the same for all values of $D$. In our fits we set $\epsilon=2$.

A preliminary study shows that the results obtained for  
$\xi_{2nd}/L=0.56404$ are more
stable than those for $Z_a/Z_p=0.19477$. In particular, the values of
$c_2$, $c_3$, ... have a smaller modulus for $\xi_{2nd}/L=0.56404$ than for 
$Z_a/Z_p=0.19477$. Therefore in the following we shall focus on $U_4$ and
$U_6$ at $\xi_{2nd}/L=0.56404$. 

\begin{figure}
\begin{center}
\includegraphics[width=14.5cm]{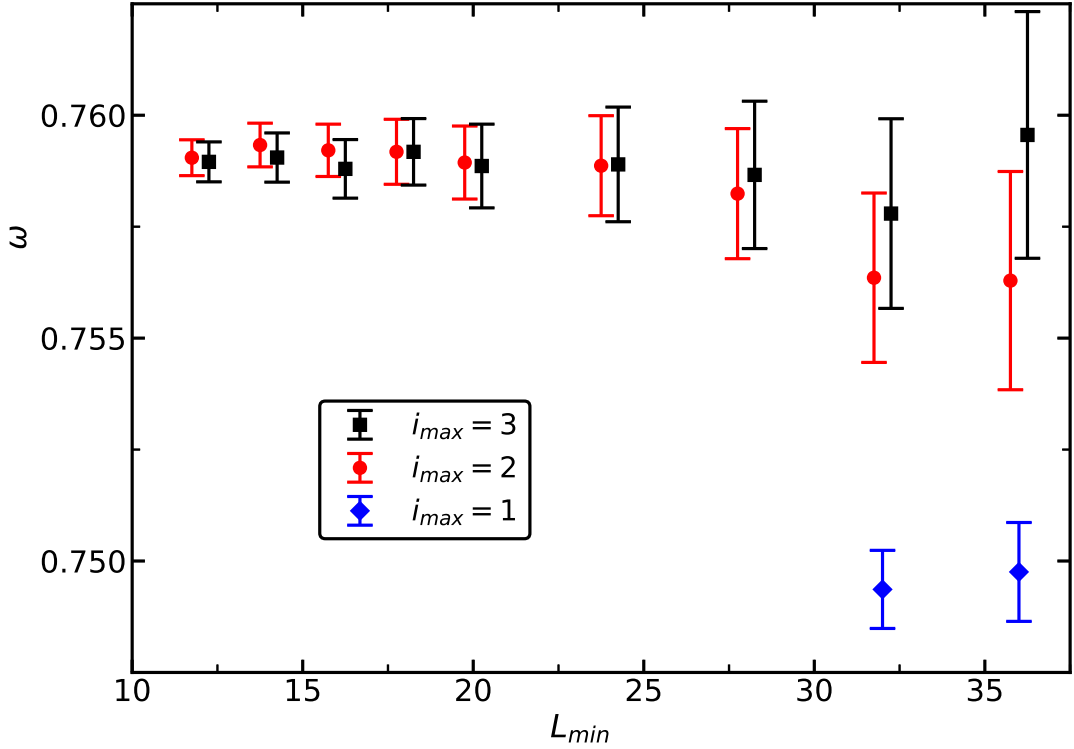}
\caption{\label{omega7}
We plot the estimates of $\omega$ obtained by fitting $U_4$ at 
$\xi_{2nd}/L=0.56404$ for $D=0.5$, $1.0$, $1.4$, $2.0$, $2.05$, $2.1$, 
and $\infty$ by using the ansatz~(\ref{manyDfit}) as a function of the 
minimal lattice size $L_{min}$ that is taken into account.
Only results for $Q > 0.01$ are given.
To make the figure more
readable, we have shifted the values of $L_{min}$ slightly.
}
\end{center}
\end{figure}

In Fig. \ref{omega7} we plot our results for the correction exponent 
$\omega$ obtained from joint fits of $U_4$ at $\xi_{2nd}/L=0.56404$ 
for $D=0.5, ..., \infty$. We give our results for 
$i_{max}=1$, $2$ and $3$ as a function of the minimal lattice size $L_{min}$
that is taken into account.
We find that fits with $i_{max}=2$ and $3$ are consistent and acceptable 
fits are obtained starting from $L_{min}=12$.  In contrast, for $i_{max}=1$
we get $Q>0.1$ only for $L_{min} \ge 40$. The estimate of $\omega$ obtained 
with $i_{max}=1$ is considerably smaller than for $i_{max}=2$ and $3$. 
As our preliminary estimate we take $\omega=0.7589(10)$ from $L_{min}=20$. 

Performing a similar analysis, taking into account all values of $D$, we get
consistent results for $\omega$ starting from $i_{max}=3$ and $4$. Here
we take $\omega=0.7595(10)$ from $L_{min}=24$ as preliminary estimate.
Taking the set $D=1.4$, ..., $\infty$ we get already consistent results for
$\omega$ with $i_{max}=1$ and $2$. Here we take $\omega=0.7584(10)$ from  
$L_{min}=20$ as preliminary estimate. 
As a check we performed fits without the term $d L^{-\epsilon}$.
We find that $\chi^2/$d.o.f. considerably increases. However the estimates for
$\omega$ are consistent with those obtained from fits with such a term.

Analyzing $U_6$ at $\xi_{2nd}/L=0.56404$ we get similar results as for $U_4$.

As our final estimate we quote
\begin{equation}
\label{omegafinal}
\omega = 0.759(2)  \;,
\end{equation}
which covers the preliminary estimates discussed above.
From the analysis of $U_4$ and $U_6$ at $Z_a/Z_p=0.19477$ we would arrive at
$\omega=0.758(4)$. Since the fits for $\xi_{2nd}/L=0.56404$ are clearly 
better behaved than those for $Z_a/Z_p=0.19477$, we stick with the result
obtained from $\xi_{2nd}/L=0.56404$ as our final estimate.

\subsection{$D^*$}
\label{Dstar}
\begin{figure}
\begin{center}
\includegraphics[width=14.5cm]{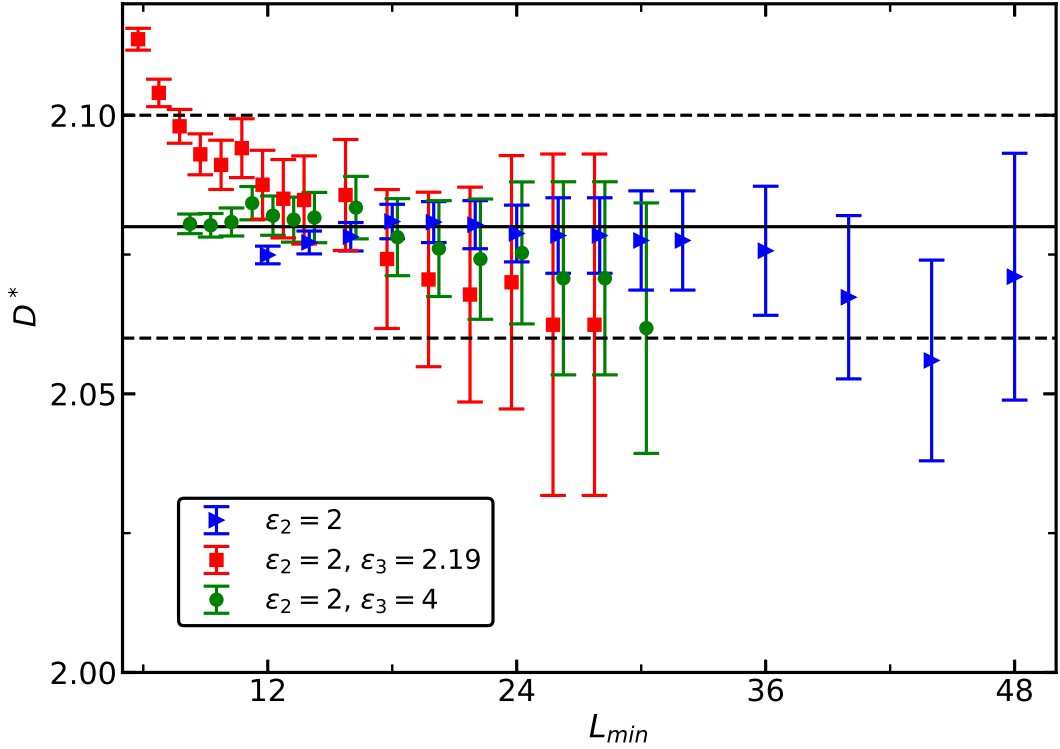}
\caption{\label{DsU4atxi}
We plot  estimates of $D^*$ obtained from fitting $U_4$ at 
$\xi_{2nd}/L=0.56404$ by using the ans\"atze~(\ref{Ubarfit1},\ref{Ubarfit2})
as a function of $L_{min}$. Only results for $Q > 0.01$ are given.
To make the figure more
readable, we have shifted the values of $L_{min}$ slightly.
The solid line indicates the preliminary results based on this set of fits.
The dashed lines give the error estimate.
}
\end{center}
\end{figure}

Here we analyze $\bar{U}_4$ and $\bar{U}_6$ at $D=2.0$, $2.05$ and $2.1$. 
To this end we consider the ans\"atze
\begin{eqnarray}
\label{Ubarfit1}
\bar{U} &=& \bar{U}^* + b(D) L^{-\epsilon_1} + c L^{-\epsilon_2} \;, \\
\label{Ubarfit2}
\bar{U} &=& \bar{U}^* + b(D) L^{-\epsilon_1} + c L^{-\epsilon_2} 
 + d L^{-\epsilon_3} \;,
\end{eqnarray}
where we fix $\epsilon_1=0.759$, which is our estimate of $\omega$ obtained 
above, $\epsilon_2=2$, which effectively takes into account $2-\eta$ and 
$\omega_{NR}$. We take either $\epsilon_3=2.19$, which corresponds to 
$\omega_{ico}$, or $\epsilon_3=4$. 
We parameterize the leading correction as $b(D) = b_{s} \; (D-D^*)$,
where $b_s$ and $D^*$ are free parameters of the fit. Furthermore $\bar{U}^*$, 
$c$ and $d$ are free parameters. Here we assume that $c$ and $d$ are the 
same for all three values of $D$, which should be a reasonable approximation.
Below we focus on $\bar{U}_4$, since the results for $\bar{U}_6$ are similar.
In Fig. \ref{DsU4atxi} we plot results obtained for $D^*$ by fitting 
$U_4$ at $\xi_{2nd}/L=0.56404$ using the ans\"atze~(\ref{Ubarfit1},\ref{Ubarfit2}).
As our preliminary estimate we take $D^*=2.08(2)$. In a similar fashion we 
arrive at $\bar{U}_4^* = 1.139295(20)$ for $\xi_{2nd}/L=0.56404$.  

Analyzing $U_4$ at $Z_a/Z_p=0.19477$  we arrive at $D^*=2.07(3)$ 
and $\bar{U}_4^* = 1.13930(2)$.

As our final estimate for the improved model we take 
\begin{equation}
 D^* = 2.08(2) \;,
\end{equation}
which is the result of the joint analysis performed in section \ref{fixedpoint}
and the analysis of $U_4$ at $\xi_{2nd}/L=0.56404$.

\subsection{The critical exponent $\nu$ of the correlation length}
\label{sectnu}

We compute the exponent $\nu=1/y_t$ from the slope of a phenomenological
coupling $R_j$ at a given
value $R_{i,f}$ of a second quantity $R_i$, where $R_j$ and $R_i$ might be 
the same. Following the discussion of section III C of ref. \cite{myClock}
these slopes behave as
\begin{eqnarray}
\label{slope}
\bar{S}_{i,j} =
\left . \frac{\partial R_j}{\partial \beta } \right |_{R_i=R_{i,f}} &=& a
L^{y_t } \; \left [1 + b L^{-\omega} + ... + c_{back} L^{-2-\eta} + c_{NR} L^{-\omega_{NR}} + c_{ico} L^{-\omega_{ico}} + ... \right] \nonumber \\
 & + & d L^{-\omega} + ...  \;.
\end{eqnarray}
Note that the coefficients $a, b, c_{back}, c_{NR}, c_{ico}$ and $d$ 
depend on the quantity that is considered and 
on the model, which means in the present case on the parameter $D$. 
As discussed in ref. \cite{myClock} and references therein it is advantageous
to take $R_{i,f} \approx R_i^*$, since otherwise an effective correction 
$\propto (R_{i,f} -R_i^*) L^{-y_t}$ has to be taken into account.

Below we consider $D=2.05$ and $2.1$ which are close to $D^*$.  Therefore 
the coefficient $b$ of the leading correction is small for all quantities.
In order to ensure that leading corrections to scaling can be safely ignored 
at the 
level of our accuracy, we construct improved slopes by multiplying 
$\bar{S}$ by a certain power $p$ of the Binder cumulant $\bar{U}_4$:
\begin{equation}
\label{improvep}
 \bar{S}_{imp} = \bar{S} \bar{U}_4^p \;,
\end{equation}
where both $\bar{S}$ and $\bar{U}_4$ are taken at $R_{i,f}$.  
The exponent $p$ is chosen
such that, at the level of our numerical accuracy, leading corrections to
scaling are eliminated. This idea is discussed systematically in
ref. \cite{ourdilute}. To determine $p$, we consider the pair 
$(D_1,D_2)=(1.4, \infty)$. Note that as discussed in section \ref{corrections},
the amplitude of the leading correction to scaling for these two values of 
$D$ has approximately the same modulus but opposite sign.
We fit ratios of $\bar{S}_{i,j}$ and $\bar{U}_4$ with the ans\"atze
\begin{equation}
\label{RSfit}
\frac{\bar{S}_{i,j} (D_1)}{\bar{S}_{i,j}(D_2)} = a_S (1+ b_S L^{- \epsilon_1} ) 
\;\;,\; \; \; \;
\frac{\bar{S}_{i,j} (D_1)}{\bar{S}_{i,j}(D_2)}  = a_S (1+ b_S L^{- \epsilon_1} + c_S L^{- \epsilon_2} )
\end{equation}
and
\begin{equation}
\label{RUfit}
\frac{\bar{U}_4(D_1)}{\bar{U}_4(D_2)} = 1 + b_U L^{- \epsilon_1} \;\;,
\;\; \;\;
\frac{\bar{U}_4(D_1)}{\bar{U}_4(D_2)} = 1 + b_U L^{- \epsilon_1} +c_U L^{- \epsilon_2}\;,
\end{equation}
where we fixed $\epsilon_1=0.76$ and $\epsilon_2=2$.  
The exponent $p$ is given by
\begin{equation}
 p = -\frac{b_S}{b_U} \;.
\end{equation}
In table \ref{pimprove} we give our final results for $p$. The error bar
takes into account statistical errors as well as systematical ones, which are 
estimated by comparing the results of the two different ans\"atze. 
\begin{table}
\caption{\sl \label{pimprove}
Numerical result for the exponent $p$ that eliminates
leading corrections to scaling in $\bar{S}_{ij}$, eq.~(\ref{improvep}).
}
\begin{center}
\begin{tabular}{ccccc}
\hline
Fixing $\backslash$  Slope of & \phantom{000} $Z_a/Z_p$  \phantom{000} &
 \phantom{000} $\xi_{2nd}/L$  \phantom{000}  &
 \phantom{0000} $U_4$  \phantom{0000} &  \phantom{0000}$U_6$  \phantom{0000}\\
\hline
$Z_a/Z_p=0.19477$:     &1.65(10)& 0.24(10) & -3.6(2) & -5.0(3) \\
$\xi_{2nd}/L=0.56404$: &0.07(7) & 0.24(10) & -4.22(10) & -5.64(10)  \\
\hline
\end{tabular}
\end{center}
\end{table}

Let us briefly comment on the statistical error of the different quantities.
We find that both fixing $Z_a/Z_p=0.19477$  and $\xi_{2nd}/L=0.56404$ changes,
compared with fixed $\beta$, the relative statistical error of the slopes 
only little. The same holds for the comparison of the improved and the 
unimproved slopes. We see big differences between the relative statistical 
errors of the slopes of the different phenomenological couplings. 
The relative error is the smallest in the case of $\xi_{2nd}/L$.  The 
ratios of the statistical errors vary only little with the linear lattice size.
For example for 
$D=2.05$ and $L=40$ at $\beta_c$ we find that the relative statistical error of 
the slope of $Z_a/Z_p$, $U_4$ and $U_6$ is by a factor of $1.24$, $1.99$, 
and $2.00$ larger than that of $\xi_{2nd}/L$.  
As a measure of the effort to reach a certain accuracy, 
beyond the increase due to the increasing lattice size, we studied
$w=n_{stat} \epsilon_r^2$, where $n_{stat}$ is the number of measurements
and $\epsilon_r$ the relative statistical error. 
We fitted the data for the slope of $Z_a/Z_p$ at $Z_a/Z_p=0.19477$ for 
$D=2.05$. We find a behavior $w \propto L^{x}$ with $x \approx 0.36$. 
Since we already averaged over bins during the simulation, we can not
determine to what extend this degradation of the efficiency is due to 
an increase of autocorrelation times or and increase of the variance of the 
slope.

Below we perform throughout joint fits of the data for $D=2.05$ and $D=2.1$. 
In these fits, the overall amplitude for each value of $D$ is a free parameter
of the fit. In contrast, we assume that the correction amplitudes are 
similar, and are taken to be same in the ansatz.

\begin{figure}
\begin{center}
\includegraphics[width=14.5cm]{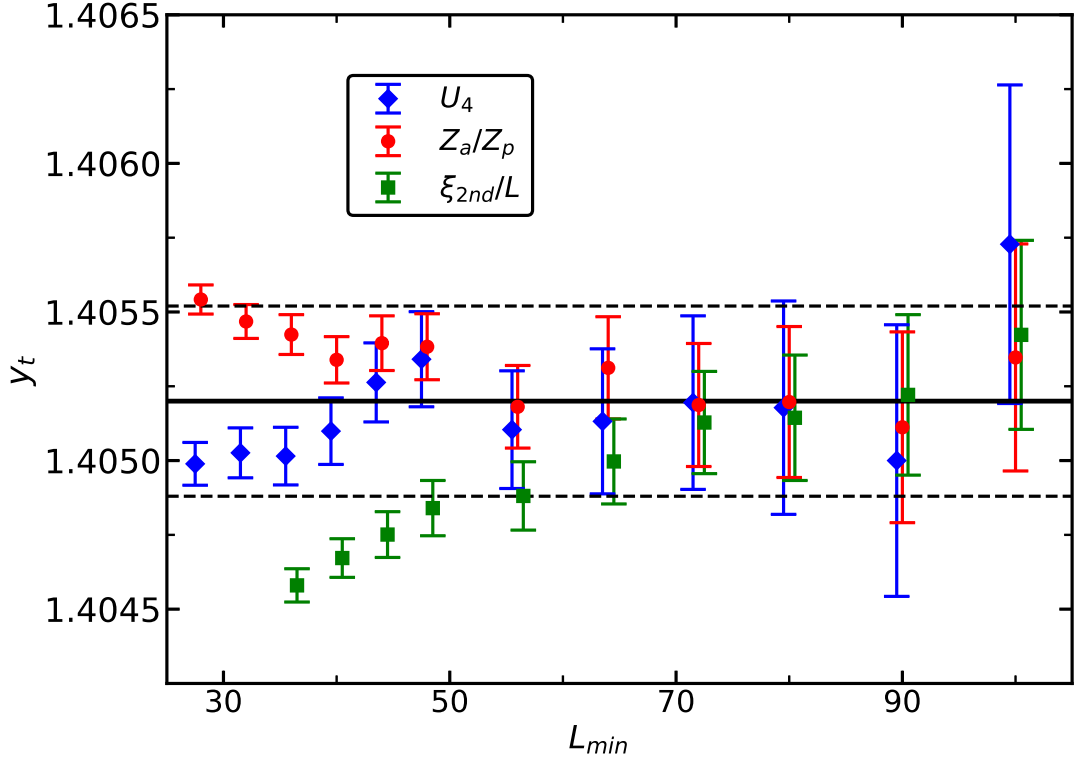}
\caption{\label{nufit0}
We plot the estimates of $y_t$ obtained from fitting the improved slopes of
$Z_a/Z_p$, $\xi_{2nd}/L$ and $U_4$ at $\xi_{2nd}/L=0.56404$ for $D=2.05$ and
$2.1$  by using the ansatz~(\ref{slope1}) as a function of the minimal 
lattice size $L_{min}$ that is taken into account.
Only results for $Q > 0.01$ are given. To make the figure more
readable, we have shifted the values of $L_{min}$ slightly.
The solid line indicates the preliminary result based on this set of fits.
The dashed lines give the error estimate.
}
\end{center}
\end{figure}

First we have analyzed the improved slopes of the different phenomenological
couplings separately. We have fitted these quantities by using the ans\"atze

\begin{eqnarray}
S &=& a L^{y_t}  \; , \label{slope1} \\
S &=& a L^{y_t} \;(1 + b L^{-\epsilon_1})  \label{slope2} \;,
\end{eqnarray}
where we take $\epsilon_1=2$, which should effectively take into account
corrections due to the analytic background of the magnetic susceptibility
and the violation of the rotational invariance by the simple cubic lattice.
First we analyzed our data by using ansatz~(\ref{slope1})
without correction term. In Fig. \ref{nufit0} we give our results
for improved slopes at $\xi_{2nd}/L=0.56404$. We do not give results for 
$U_6$, since they are very similar to those for $U_4$. 
Note that for example for $L_{min}=40$ we get $\chi^2/$d.o.f.$=0.949$ and
$1.150$ for the slopes of $\xi_{2nd}/L$ and $Z_a/Z_p$, respectively.
This corresponds to $Q=0.526$ and $0.286$, respectively. Despite this fact, 
the estimates of $y_t$ obtained for $L_{min}=40$ clearly differ for 
$\xi_{2nd}/L$ and $Z_a/Z_p$. 
As our preliminary estimate we take $y_t = 1.40520(32)$. It is chosen such that
all three results for $L_{min}=72$ are covered.
The estimates obtained for $Z_a/Z_p=0.19477$ are similar.

\begin{figure}
\begin{center}
\includegraphics[width=14.5cm]{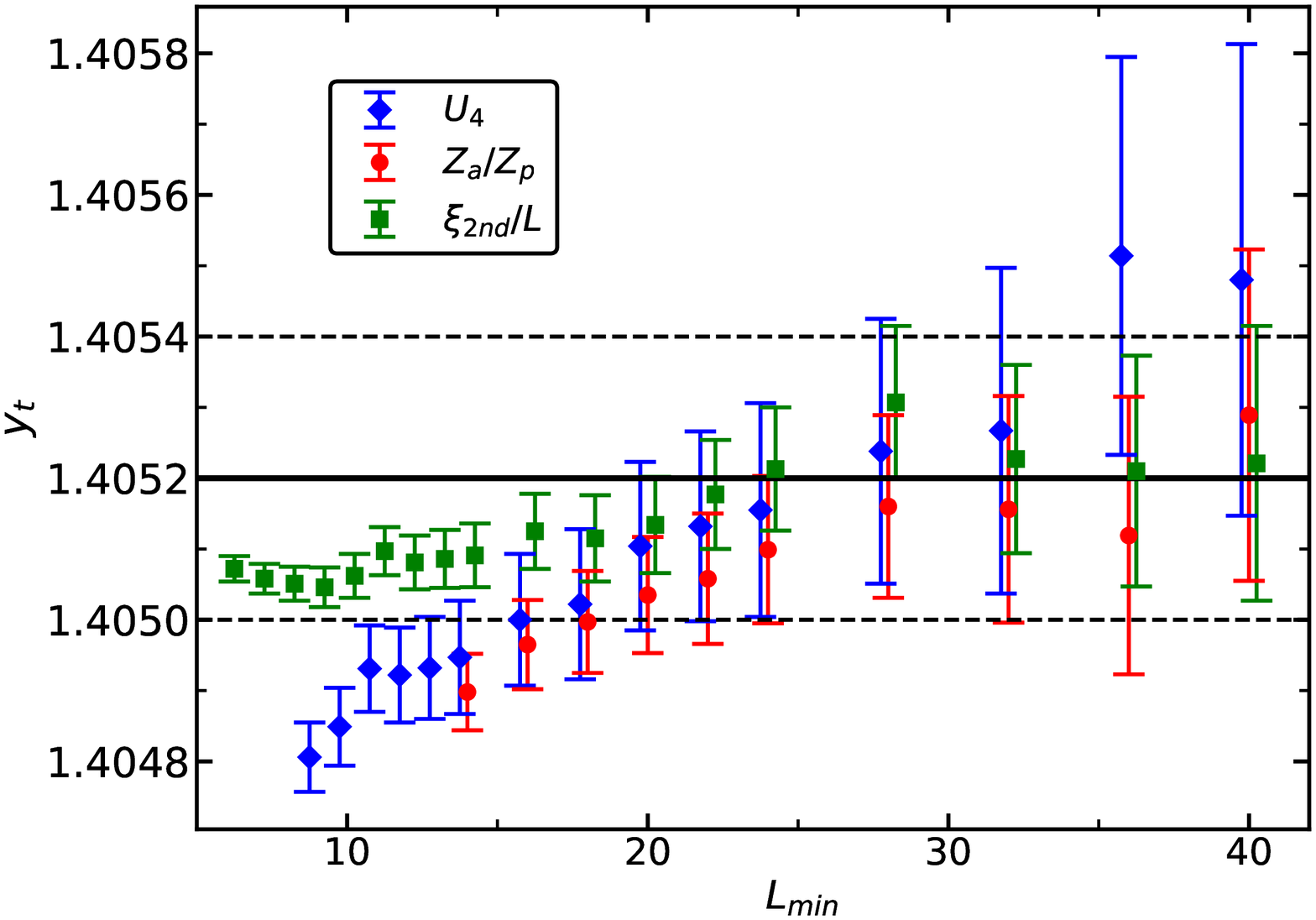}
\caption{\label{nufit1}
We plot the estimates of $y_t$ obtained from fitting the improved slopes of 
$Z_a/Z_p$, $\xi_{2nd}/L$ and $U_4$ at $\xi_{2nd}/L=0.56404$ for $D=2.05$ and 
$2.1$ by using the ansatz~(\ref{slope2}) as a function of the minimal lattice 
size $L_{min}$ that is taken into account. Only results for $Q > 0.01$ are 
shown. To make the figure more readable, we have shifted the values of 
$L_{min}$ slightly. The solid line indicates our preliminary estimate of 
this set of fits. The dashed lines give the error estimate.
}
\end{center}
\end{figure}

In Fig. \ref{nufit1} we give results obtained from fitting the improved 
slopes of $Z_a/Z_p$, $\xi_{2nd}/L$ and $U_4$ at $\xi_{2nd}/L=0.56404$ by using
the ansatz~(\ref{slope2}).  As our preliminary estimate of this set of fits
we take $y_t = 1.40520(20)$. It covers all three estimates obtained 
for $L_{min}=24$. 
Analyzing the slopes at $Z_a/Z_p=0.19477$ in a similar way, we find 
consistent results. 

\begin{figure}
\begin{center}
\includegraphics[width=14.5cm]{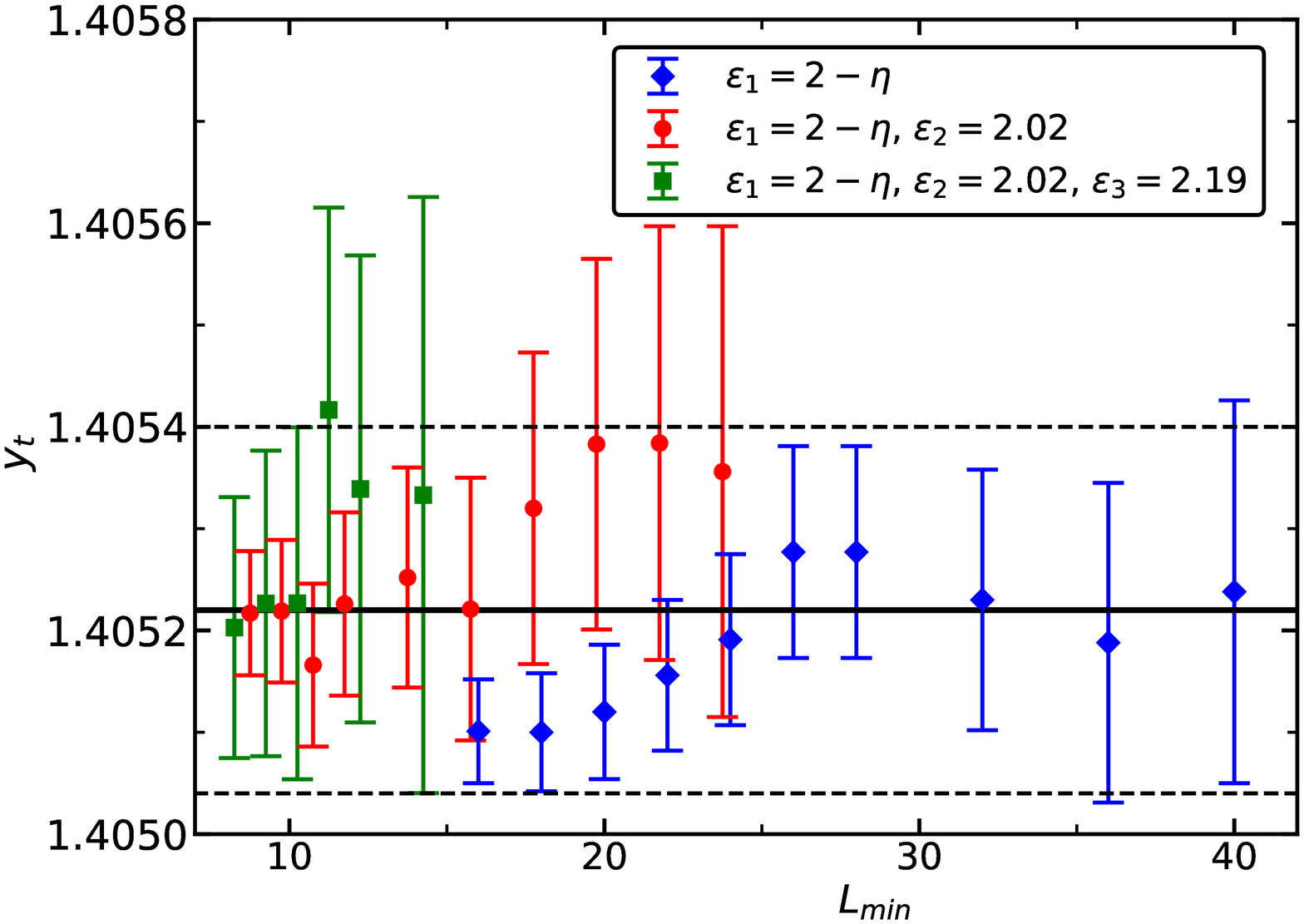}
\caption{\label{nufit2}
We plot the estimates of $y_t$ obtained by fitting the improved slopes of
$Z_a/Z_p$, $\xi_{2nd}/L$, $U_4$ and $U_6$ at $\xi_{2nd}/L=0.56404$ 
jointly by using up to three different correction terms 
as a function of the minimal lattice size $L_{min}$
that is taken into account. Data for $D=2.05$ and $D=2.1$ are taken.
Only results for $Q > 0.01$ are given. To make the figure more
readable, we have shifted the values of $L_{min}$ slightly.
The solid line indicates our preliminary estimate based on this set of fits.
The dashed lines give the error estimate.
The exponents $\epsilon_i$ given in the legend refer to the correction terms
that are included in the ans\"atze.
}
\end{center}
\end{figure}

\begin{figure}
\begin{center}
\includegraphics[width=14.5cm]{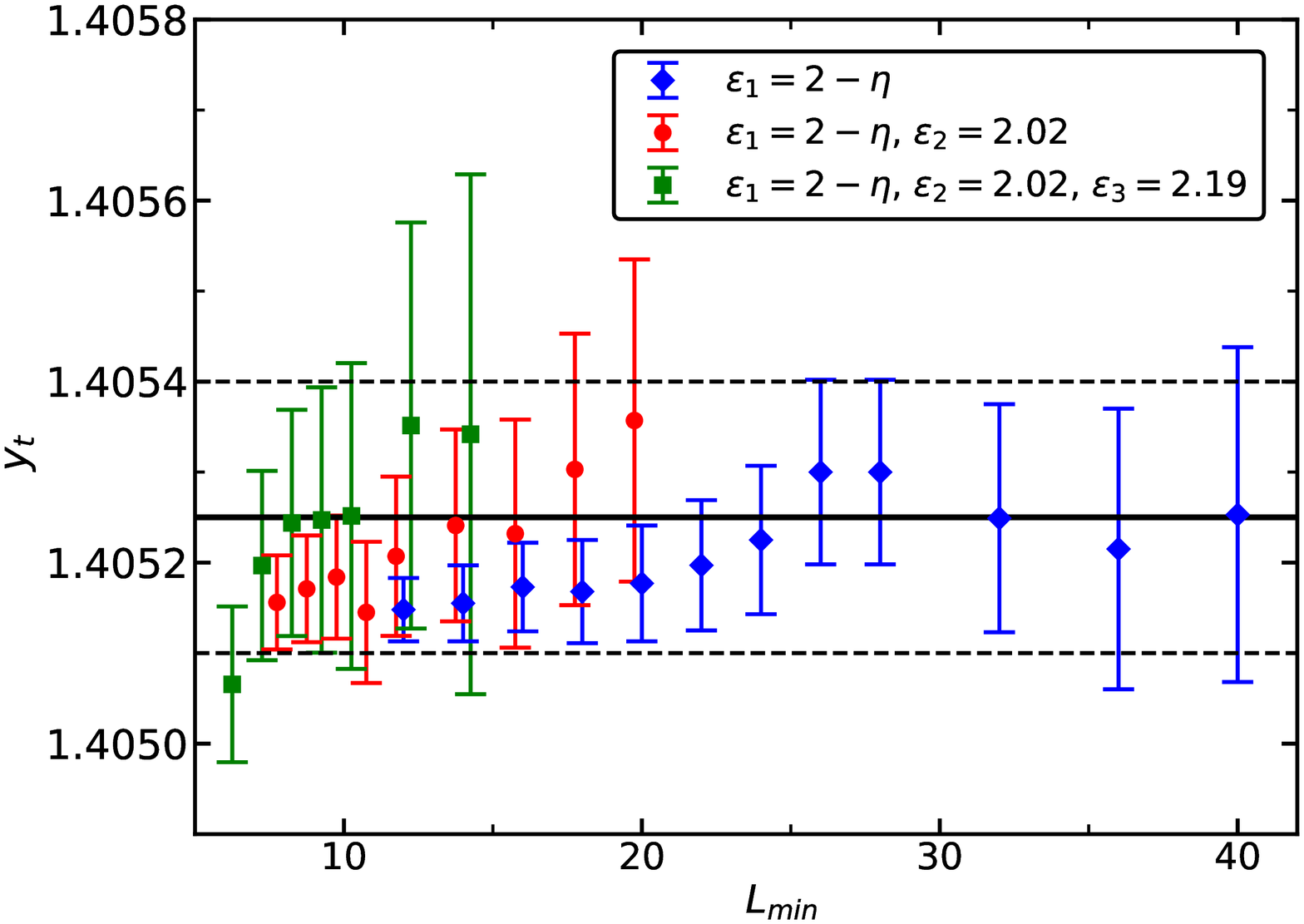}
\caption{\label{nufit3}
Same as Fig. \ref{nufit2} but for $Z_a/Z_p=0.19477$ instead of  
$\xi_{2nd}/L=0.56404$.
}
\end{center}
\end{figure}

Finally we performed a joint analysis of the improved slopes of all four 
phenomenological couplings at either $\xi_{2nd}/L=0.56404$
or $Z_a/Z_p=0.19477$. Similar to section \ref{fixedpoint}, 
we took the covariances of the different quantities into account.
In these fits, we used ans\"atze with up to three different correction
terms with the effective correction exponents $\epsilon_1 = 2-\eta$, 
$\epsilon_2 = 2.02$, and $\epsilon_3 = 2.19$. The first is motivated by
the analytic background of the magnetic susceptibility, the second by
$\omega_{NR}$ and the third by $\omega_{ico}$. Note that in the slope we 
also expect corrections with the exponent $y_t + \omega$, which is effectively
taken into account by $\epsilon_3$. 
In Fig. \ref{nufit2} we give our results for the improved slopes at
$\xi_{2nd}/L=0.56404$. In Fig. \ref{nufit3} we give the corresponding 
results for $Z_a/Z_p=0.19477$.  As our preliminary estimates we take 
$y_t = 1.40522(18)$ and $1.40525(15)$ obtained for  $\xi_{2nd}/L=0.56404$ 
and $Z_a/Z_p=0.19477$, respectively.

Based on these results and the preliminary estimate obtained by fitting 
the slopes of the different phenomenological couplings separately by
using the ansatz~(\ref{slope2}) we conclude
\begin{equation}
y_t = 1.4052(2) \;,
\end{equation}
which corresponds to $\nu=0.71164(10)$. 

\subsection{The critical exponent $\eta$}
\label{secteta}
We analyzed the improved quantities
\begin{equation}
\label{chiimp}
\bar{\chi}_{imp} = \bar{\chi}  \bar{U}_4^p \;,
\end{equation}
where both $\chi$ and $U_4$ are taken either at $Z_a/Z_p =0.19477$ or
$\xi_{2nd}/L =0.56404$. We computed the exponent $p$ in a similar way as 
in the previous section for the slopes $S$. Therefore we skip a detailed 
discussion and only report our results $p=-1.31(3)$ and $-0.23(4)$ for
$Z_a/Z_p =0.19477$ and $\xi_{2nd}/L =0.56404$, respectively.

Let us briefly discuss the effect of taking $\chi$ at $Z_a/Z_p =0.19477$
or $\xi_{2nd}/L =0.56404$ on the statistical error. In previous work, see  
ref. \cite{myClock} and references therein,  we observed that the statistical 
error is reduced compared with $\chi$ at a fixed value of 
$\beta \approx \beta_c$.
Here we see for $Z_a/Z_p =0.19477$  only a small effect, 
while for $\xi_{2nd}/L =0.56404$ we see for example for $D=2.05$ a reduction 
of the statistical error by a factor of about two. 
The relative statistical error of the improved susceptibility is 
by a few percent larger than that of the unimproved counterpart. 

\begin{figure}
\begin{center}
\includegraphics[width=14.5cm]{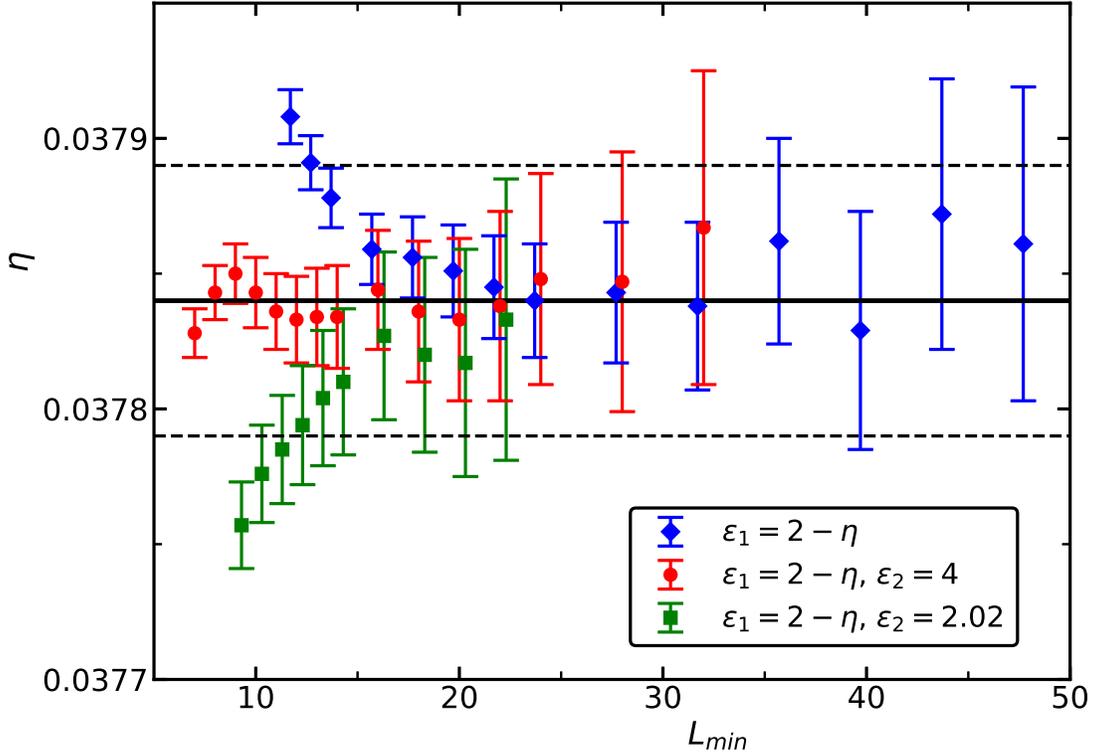}
\caption{\label{etaZI}
We plot the estimates of $\eta$ obtained from fitting the improved
magnetic susceptibility at $Z_a/Z_p =0.19477$ by using the 
ans\"atze~(\ref{chi1},\ref{chi2}) as a function of the 
minimal lattice size $L_{min}$ that is taken into account.
Data for $D=2.05$ and $D=2.1$ are used in the fits.
Only results for $Q >0.01$ are given. The numbers given in the legend 
refer to the corrections that are taken into account in the ansatz.
To make the figure more
readable, we have shifted the values of $L_{min}$ slightly.
The solid line gives our preliminary estimate and the dashed lines 
indicate the error bar.
}
\end{center}
\end{figure}

We fitted our data with the ans\"atze 
\begin{eqnarray}
\label{chi0}
\chi &=& a L^{2-\eta} \;, \\ 
\label{chi1}
\chi &=& a L^{2-\eta} + b \;, \\
\label{chi2}
\chi &=& a L^{2-\eta} \;(1 + c L^{-\epsilon_2}) \; + b  \;,
\end{eqnarray}
where the analytic background $b$ can be viewed as an effective correction 
with the exponent $\epsilon_1=2-\eta$. Similar to the analysis of the 
slopes, we performed joint fits of the data for $D=2.05$ and $2.1$. 

In Fig. \ref{etaZI} we give the estimates obtained from fitting 
the improved magnetic susceptibility at $Z_a/Z_p =0.19477$ by using the
ans\"atze~(\ref{chi1},\ref{chi2}). In the case of ansatz~(\ref{chi2})
we plot results for $\epsilon_2=2.02$ and $4$. We also performed 
fits using $\epsilon_2=2.19$, which give consistent results for $\eta$.
Our preliminary estimate $\eta = 0.03784(5)$ for this set of fits
is consistent with the estimate obtained by using ansatz~(\ref{chi2}) with
$\epsilon_2=2.02$ for $L_{min}=16$. Furthermore it covers the results
obtained by using the ansatz~(\ref{chi1}) for $L_{min}=14$ up to $32$
and ansatz~(\ref{chi2}) with $\epsilon_2=4$ for $L_{min} \le 24$.
Fitting the improved magnetic susceptibility at $\xi_{2nd}/L =0.56404$
by using the ans\"atze~(\ref{chi1},\ref{chi2}) we find results that are
consistent with the estimate $\eta = 0.03784(5)$.

\begin{figure}
\begin{center}
\includegraphics[width=14.5cm]{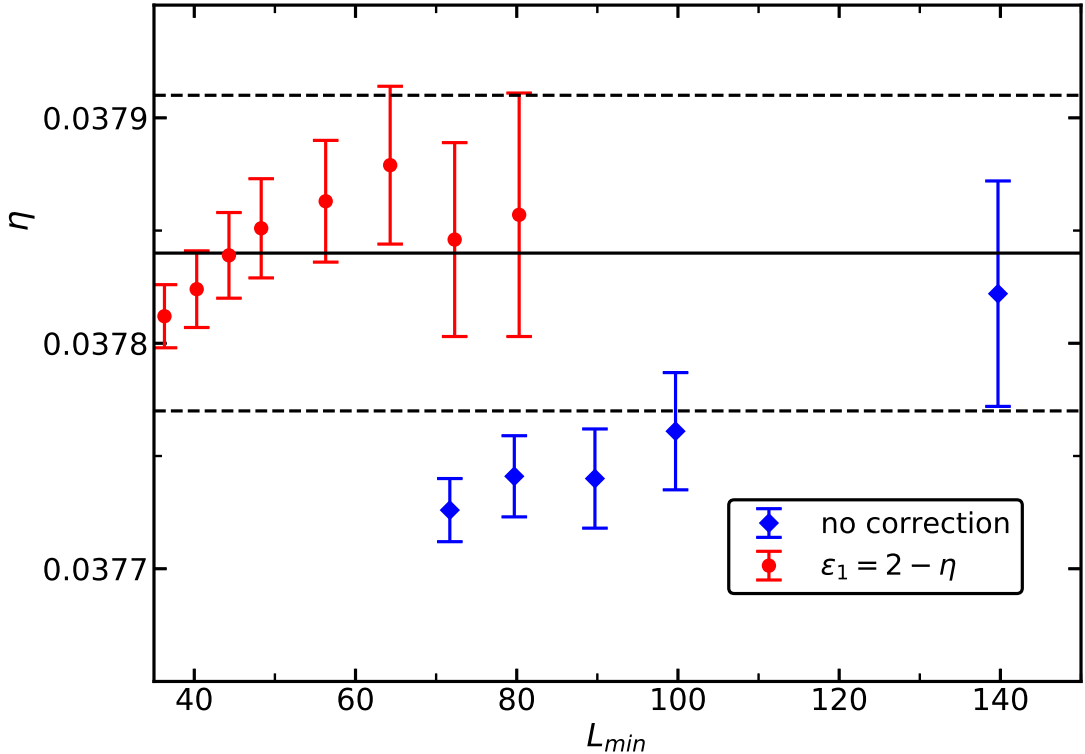}
\caption{\label{etaXI2}
We plot the estimates of $\eta$ obtained from fitting the improved
magnetic susceptibility at $\xi_{2nd}/L =0.56404$ as a function of the 
minimal lattice size $L_{min}$ that is taken into account. Data for 
$D=2.05$ and $D=2.1$ are used.
To make the figure more
readable, we have shifted the values of $L_{min}$ slightly.
Only results for $Q >0.01$ are given. Either no correction 
or a term corresponding to the analytic background is used in the ansatz.
The solid line gives our preliminary estimate of $\eta$ and the dashed lines
indicate the error bar.
}
\end{center}
\end{figure}

Finally, in Fig. \ref{etaXI2} we plot the estimates obtained from
fits of the data for the improved magnetic susceptibility at 
$\xi_{2nd}/L =0.56404$
without correction term~(\ref{chi0}) and with a correction corresponding to
the analytic background of the magnetic susceptibility, eq.~(\ref{chi1}).
Based on these fits we arrive at the preliminary estimate $\eta=0.03784(7)$ 
which takes into account the result obtained by using the ansatz~(\ref{chi0})
with $L_{min}=140$ and the results obtained by using the ansatz~(\ref{chi1}) 
up to $L_{min}=80$. As our final estimate of $\eta$ we quote 
\begin{equation}
\eta=0.03784(5)
\end{equation}
obtained by using ans\"atze that include correction terms.
\section{Summary and discussion}
\label{summary}
We have studied the generalized icosahedral model on the simple 
cubic lattice. In this model, the field variable might take a normalized 
vertex of the icosahedron as value. Analogous to the Blume-Capel model, 
in addition $(0,0,0)$ is allowed.  The density of the $(0,0,0)$ 
sites is governed by the parameter $D$ of the reduced Hamiltonian.
For a certain range of $D$, the model undergoes a second-order phase
transition. At the critical line, the symmetry is enhanced to $O(3)$.
Hence the transition belongs to the universality class of the 
three-dimensional Heisenberg model. In the Appendix \ref{omegaico} we find 
that a perturbation of the $O(3)$-invariant fixed point with the symmetry
of the icosahedron is related with the irrelevant RG-eigenvalue 
$y_{ico} = - 2.19(2)$. On the critical line, the amplitude of leading 
corrections to scaling depends of the parameter $D$. Numerically we 
find that for $D^* = 2.08(2)$ this amplitude vanishes. Based on a finite
size scaling analysis of phenomenological couplings, such as the Binder 
cumulant, their slopes and the magnetic susceptibility we arrive at 
accurate estimates of the critical exponents $\nu$ and $\eta$ and the correction
exponent $\omega$.
In Appendix \ref{appendixA} we analyze data obtained for the 
three-component $\phi^4$ model on the simple cubic lattice, leading 
to consistent results for the exponents $\nu$ and $\eta$, confirming
that both models share the same universality class. The precision of 
our results clearly surpasses that of experiments. However one should
note that there had been theoretical advances in recent years made
by different methods. Here our results serve as benchmark. 
In the introduction, in table \ref{methods} we confront our results
with ones given in the literature. Comparing with 
the $\epsilon$-expansion, we find significant deviations. The estimates
obtained by using the conformal bootstrap method \cite{Kos:2016ysd} 
and the recent implementation of the functional renormalization group
method \cite{DePo20} are consistent with but less precise than ours.

Our precise estimates of the inverse critical temperature for various
values of $D$ and $\lambda$ for the generalized icosahedral model and 
the $\phi^4$ model, respectively, might serve as input for studies 
focussing on other properties of these models. In particular
we intend to compute the structure constants using a similar approach as
in ref. \cite{myStructure} for the Ising universality class.
Furthermore it would be interesting to investigate the symmetry properties
of the icosahedral model in the low-temperature phase.

Our motivation to study the icosahedral model is of technical nature.
In order to save the field variable at one site only 4 bits are needed.
For practical reasons, in our program a 8 bit \verb+char+ variable is 
used. Furthermore, probabilities needed for the Metropolis and the cluster 
update can be computed and tabulated at the beginning of the simulation.
For our implementation we find a speed up by roughly a factor of 
three compared with the $\phi^4$ model studied for example in refs.
\cite{myO3O4,ourHeisen,HaVi11}.  This advantage is partially abrogated by
the correction $\propto L^{y_{ico}}$ that is not present in a model with
$O(3)$ symmetry at the microscopic level. 
Note that the situation is different for the $(q+1)$-state clock model 
studied in ref. \cite{myClock}. In this case the irrelevant exponent $y_q$
is rapidly decreasing with $q$. In ref. \cite{myClock} we focused on 
$q=8$, where $y_{q=8} = -5.278(9)$, see ref. \cite{Debasish2}. 
Hence the correction can be ignored in the analysis of the data, meaning
that we have the technical advantage without a downside.

\section{Acknowledgement}
This work was supported by the Deutsche Forschungsgemeinschaft (DFG) under 
grant No HA 3150/5-1.

\appendix

\section{The $\phi^4$ model on the lattice}
\label{appendixA}
The $\phi^4$ model on the simple cubic lattice is defined by the reduced
Hamiltonian
\begin{equation}
 {\cal H}_{\phi^4}= -\beta \sum_{<xy>} \vec{\phi}_x \cdot  \vec{\phi}_y 
+ \sum_x \left [ \vec{\phi}_x^{\,2} + \lambda (\vec{\phi}_x^{\,2} -1)^2  \right]
\;,
\end{equation}
where $\vec{\phi}_x \in \mathbb{R}^N$ with $N=3$ in our case.   
We performed simulations for
$\lambda =5$ and $5.2$. Note that $\lambda^*=5.2(4)$, eq.~(B13) of 
ref. \cite{HaVi11}.
We simulated at $\beta=0.6875638$  and $0.687985$ in the
case of $\lambda = 5$ and $5.2$, respectively. These are the estimates of 
$\beta_c$ obtained in ref. \cite{ourHeisen} and in preliminary simulations, 
respectively. The simulations are organized in a similar fashion as for the
generalized icosahedral model. For $\lambda = 5.2$ we 
have simulated the linear lattice sizes 
$L=8$, $9$, ..., $20$, $22$, ..., $30$, $34$, $40$, $50$, $60$, $80$, $100$, 
$140$, $200$, and $300$. The number of measurements decreases with increasing 
lattice size. Up to $L=19$ we performed about $3 \times 10^9$ measurements.
For $L=300$ we performed $3.75 \times 10^7$ measurements. In total we spent
about $13.5$ years of CPU time on these simulations.
In the case of $\lambda = 5.0$ we performed simulations for fewer lattice 
sizes. We simulated at $L=8$, $10$, ..., $30$, $34$, $40$, $50$, $60$, $80$,
$100$ and $140$. In total we spent about $5.5$ years of CPU time on 
these simulations.

\subsubsection{The inverse critical temperature}
First we determine the inverse critical temperature by analyzing the 
improved phenomenological coupling $Z_a/Z_p+0.575 \; U_4$. We fit
our data with the ansatz
\begin{equation}
 R(L,\beta_c) = R^* +c L^{-2} \;,
\end{equation}
where we use the estimate $(Z_a/Z_p+0.575 \; U_4)^*=0.84987(3)$
that we obtained from the analysis
of the data for the icosahedral model in section \ref{fixedpoint}. 
As final result we get
\begin{eqnarray}
\beta_c(\lambda=5.0) &=& 0.68756127(13)[6] \; , \\
\beta_c(\lambda=5.2) &=& 0.68798521(8)[3] \; , 
\end{eqnarray}
where the error in $[]$ is due to the uncertainty of  $(Z_a/Z_p+0.575 \; U_4)^*$.

\subsubsection{The improved model}
Here we study the behavior of $U_4$ at $\xi_{2nd}/L=0.56404$ or
$Z_a/Z_p=0.19477$.  

We perform fits similar to those performed in section \ref{Dstar}. Here
we only include the data obtained for $\lambda=5.0$ and $5.2$. Furthermore
we fix the value of $\bar{U}_4^*$ to that obtained in section \ref{Dstar}.
For $\xi_{2nd}/L=0.56404$ we get $\lambda^*=5.19(2)[6]$,
while for $Z_a/Z_p=0.19477$ we get $\lambda^*=5.14(2)[6]$,
where the error in $[]$ is due to the uncertainty of $\bar{U}_4^*$.
Our final estimate
\begin{equation}
 \lambda^*= 5.17(11) 
\end{equation}
is chosen such that both estimates, including their errors are covered.

\subsubsection{Finite size scaling estimate of $\nu$}
Here we performed an analysis similar to that for the icosahedral 
model in section \ref{sectnu}. 
The data for $D=2.05$ and $2.1$ for the icosahedral
model are replaced by those for $\lambda=5.0$ and $5.2$.  Below we discuss
results obtained by analyzing improved slopes at $Z_a/Z_p=0.19477$. 
The corresponding results for $\xi_{2nd}/L=0.56404$  differ only by 
little.

\begin{figure}
\begin{center}
\includegraphics[width=14.5cm]{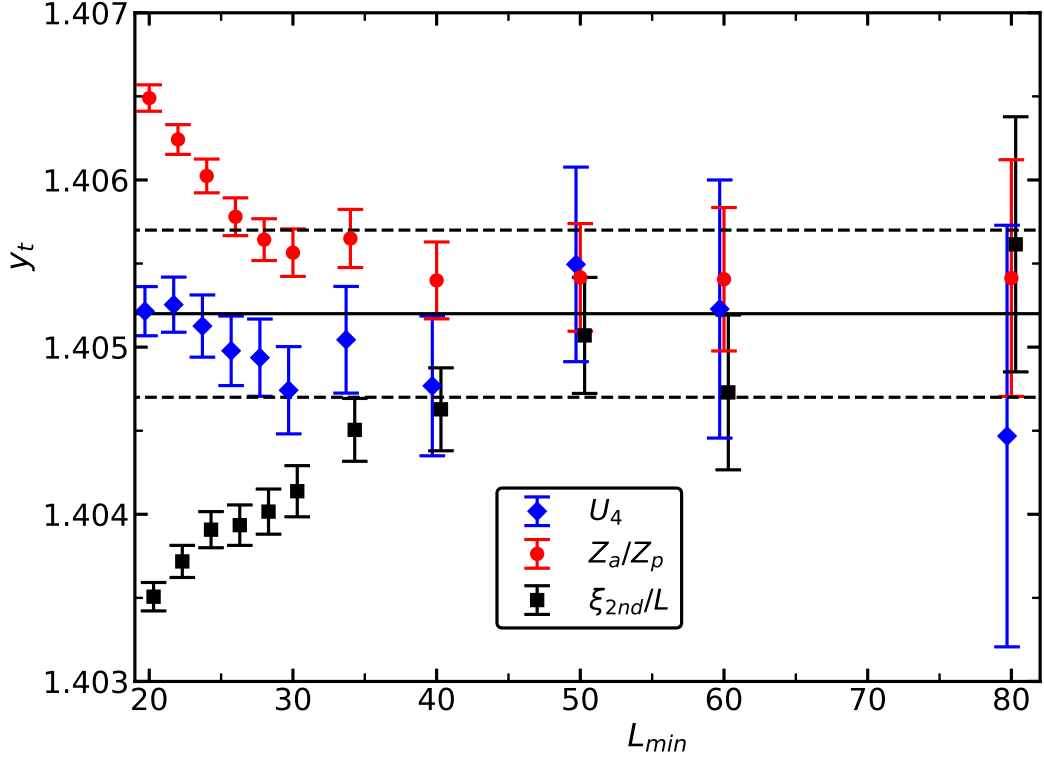}
\caption{\label{Phi4nuNoC}
We plot the estimates of $y_t$ obtained by fitting the improved slopes of 
phenomenological couplings at $Z_a/Z_p=0.19477$ 
by using the ansatz~(\ref{slope1}) without 
corrections as a function of $L_{min}$. Data for the $\phi^4$ model
at $\lambda=5$ and $\lambda=5.2$ are analysed jointly.
Only results for $Q > 0.01$ are given. To make the figure more
readable, we have shifted the values of $L_{min}$ slightly.
The solid line indicates the
final estimate that we obtain from this set of fits. The dashed
lines give the error bar.
}
\end{center}
\end{figure}

In Fig. \ref{Phi4nuNoC} we give estimates of $y_t$ obtained by using 
an ansatz without correction term, eq.~(\ref{slope1}).  Similar to 
Fig. \ref{nufit0} we find that for small $L_{min}$ the estimates obtained 
from different phenomenological couplings do not agree within their respective
error bars.
As our final estimate we take
\begin{equation}
y_t=1.4052(5) \;,
\end{equation}
corresponding to $\nu=0.71164(25)$. This estimate is consistent with results 
obtained for some range of $L_{min}$ for each of the three phenomenological 
couplings.

\subsubsection{Finite size scaling estimate of the exponent $\eta$}
\begin{figure}
\begin{center}
\includegraphics[width=14.5cm]{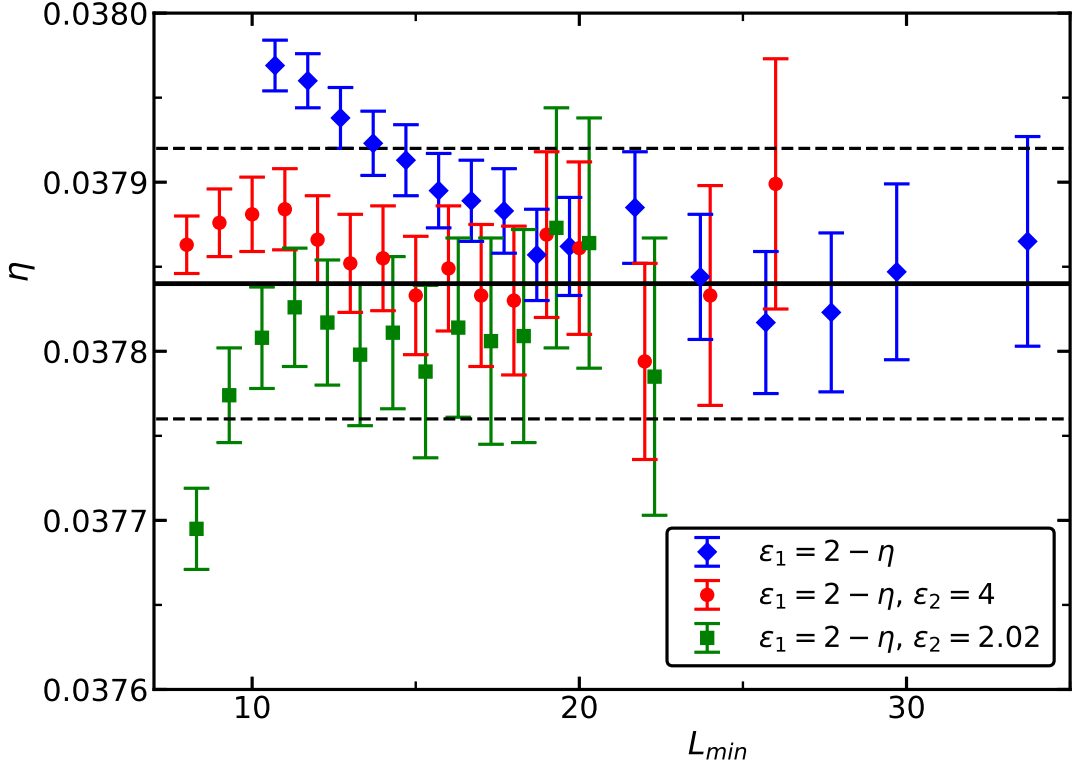}
\caption{\label{Phi4Chi}
We plot the estimates of $\eta$ obtained by fitting the improved 
magnetic susceptibility $\chi$ at $Z_a/Z_p=0.19477$ for $\lambda=5.0$ 
and $5.2$ jointly. The effective correction exponents given in the legend refer
to the ans\"atze~(\ref{chi1},\ref{chi2}). In the fits all lattice sizes
$L \ge L_{min}$ are taken into account. To make the figure more
readable, we have shifted the values of $L_{min}$ slightly.
The solid line gives the preliminary estimate that we obtain from this set 
of fits. The dashed lines give the error bar.
}
\end{center}
\end{figure}

We performed joint fits of the data for the magnetic susceptibility at
$\lambda=5.0$ and $5.2$ by using the ansatz~(\ref{chi1}) or
the ansatz~(\ref{chi2}) using either $\epsilon_2=2.02$ or $4$. We 
analyzed both the improved magnetic susceptibility at $Z_a/Z_p=0.19477$
and $\xi_{2nd}/L=0.56404$.  The results of such fits for $Z_a/Z_p=0.19477$
are plotted in Fig.~\ref{Phi4Chi}. 
Our preliminary estimate $\eta=0.03784(8)$ is chosen such that the estimates
of $\eta$ obtained by using the three different ans\"atze are contained 
in the range $0.03784 \pm 0.00008$ for some range of the minimal lattice size
$L_{min}$.  Performing a similar analysis for  $\xi_{2nd}/L=0.56404$ we arrive
at the slightly smaller estimate $\eta=0.03780(8)$. As the final estimate 
we quote
\begin{equation}
 \eta = 0.03782(10)  \;,
\end{equation}
which covers both the estimates obtained from the data for fixing 
$Z_a/Z_p=0.19477$ and $\xi_{2nd}/L=0.56404$. 

\subsubsection{Reanalysis of the high temperature series expansion}
Our more precise estimates of $\lambda^*$ and the more accurate 
estimate of $\beta_c$ at $\lambda=5.0$  are used 
to bias the analysis of the high temperature series performed in ref. 
\cite{ourHeisen}. We start from tables XXV and XXVI in appendix B of ref.
\cite{ourHeisen}. For $\lambda=4$, $4.5$ and $5$ the high temperature 
series of the magnetic susceptibility and the second moment correlation 
length is analyzed by using biased integral approximants  \cite{Gu}.
To this end, the estimates of $\beta_c$ given in eqs.~(B5,B6,B7) of \cite{ourHeisen}
are used. Let us discuss the details of our reanalysis at the example
of the exponent $\nu$, given in table XXVI of  appendix B.  The estimates 
of $\nu$  carry two types of error estimates: the number given $()$ is 
obtained from the spread of different approximants, while the number given in $[]$ 
is due to the uncertainty of the estimate of $\beta_c$.  It was obtained by 
reanalyzing the series for $\beta_c \pm \Delta \beta_c$, where $\Delta \beta_c$ is 
the estimate of the error of $\beta_c$.  Note that the value obtained for the 
exponent is increasing with an increasing estimate of $\beta_c$.  
Hence the new estimate of the exponent is given by
\begin{equation}
 \nu_{new} = \nu_{old} + (\beta_{c,new} - \beta_{c,old}) 
\frac{\Delta \nu}{\Delta \beta_c} \;.
\end{equation}
Here "new" refers to the present work, while "old" refers to 
ref. \cite{ourHeisen}.
$\Delta \beta_c$ is the error estimate of $\beta_c$ in ref. \cite{ourHeisen} 
and $\Delta \nu$ refers to the number given in $[]$ in table XXVI of 
ref. \cite{ourHeisen}.

Shifting the estimates of $\nu$ for $\lambda=5$ for the approximants
bIA1 and bIA2 we arrive at $\nu=0.71141(5)[1]$ and $0.71144(6)[1]$, 
respectively. For $\lambda=4.5$, using the estimate of $\beta_c$ obtained
in ref. \cite{HaVi11}, we arrive at $\nu=0.71103(3)[4]$ and $0.71102(6)[4]$. 
Finally, extrapolating to $\lambda^*=5.17(11)$ we arrive at 
\begin{equation}
\nu=0.7116(2)  \;.
\end{equation}
Performing a similar analysis for the exponent of the magnetic susceptibility
we arrive at $\gamma=1.3965(3)$. Note that in ref. \cite{ourHeisen} 
$\gamma=1.3960(9)$ is quoted.
Plugging in $\lambda^*=5.17(11)$ into eq.~(19) of ref. \cite{ourHeisen}
\begin{equation}
\eta \nu = 0.02665(18) + 0.00035 \; (\lambda-4.5) \;
\end{equation}
we arrive at
\begin{equation}
\eta =0.0378(3)  \;,
\end{equation}
where we took into account the uncertainty of $\lambda^*$ and $\nu$. 

\section{The correction exponent $\omega_{ico}$}
\label{omegaico}
We consider the quantity
\begin{equation}
q = \frac{\langle \mbox{max}_j  \;  \vec{v}_j \cdot \vec{m}  \rangle}
                               {\langle | \vec{m} | \rangle}  \;,
\end{equation}
where $\vec{m} = \sum_x \vec{s}_x$ is the magnetization of a given
configuration, and $\vec{v}_j$, with $j=1,2,...,12$ are the twelve
possible values of the spin with unit length.
Alternatively, one might define a quantity based of the polynomial given
in ref. \cite{Cara}. 
For an $O(3)$-invariant distribution of $\vec{m}$ the value of $q$ can be
easily computed by using numerical integration.  We get
\begin{equation}
q^* = 0.915874306174 \;.
\end{equation}
The deviation from an $O(3)$-invariant distribution is now quantified by
\begin{equation}
 \bar{q} = q - q^*  \;.
\end{equation}
We computed $q$ at either $Z_a/Z_p=0.19477$ or $\xi_{2nd}/L=0.56404$.
It turns out that the numbers for $Z_a/Z_p=0.19477$ and $\xi_{2nd}/L=0.56404$
are very similar and the estimates of $\omega_{ico}$ are essentially
the same for these two cases. Therefore we restrict the discussion below on
$Z_a/Z_p=0.19477$.
We fitted $\bar{q}$ by using the ans\"atze
\begin{equation}
\label{icofit1}
 \bar{q} = a L^{-\omega_{ico}}
\end{equation}
and 
\begin{equation}
\label{icofit2}
 \bar{q} = a L^{-\omega_{ico}} \;  (1 + c L^{-2}) \;.
\end{equation}
First we checked the effect of leading corrections to scaling. To this
end we fitted our data for $D=\infty$, $1.4$, and $1.0$ using the
ansatz ~(\ref{icofit1}) and $L_{min}=12$ in all three cases. We get
$\omega_{ico}=2.237(13)$, $2.125(7)$, and $2.069(5)$ and
$\chi^2/$d.o.f.$=0.91$, $0.78$, and $2.22$ for $D=\infty$, $1.4$, and $1.0$,
respectively. We see a clear dependence
of the result for $\omega_{ico}$ on $D$. Note that for both $D=\infty$ and
$1.4$ we get an acceptable $\chi^2/$d.o.f., while the estimates of
$\omega_{ico}$ are inconsistent. 

Based on fits with the ansatz~(\ref{manyDfit}) discussed in section 
\ref{corrections} we know that the modulus of the amplitude
of leading corrections to scaling at $D=2.05$ and $2.1$ is by about a
factor of 30 smaller than for $D=\infty$ or $1.4$. Therefore the 
effect on the estimate of $\omega_{ico}$ should roughly be given by 
$[2.237(13)-2.125(7)]/60$, which we might ignore in the following.

\begin{figure}
\begin{center}
\includegraphics[width=14.5cm]{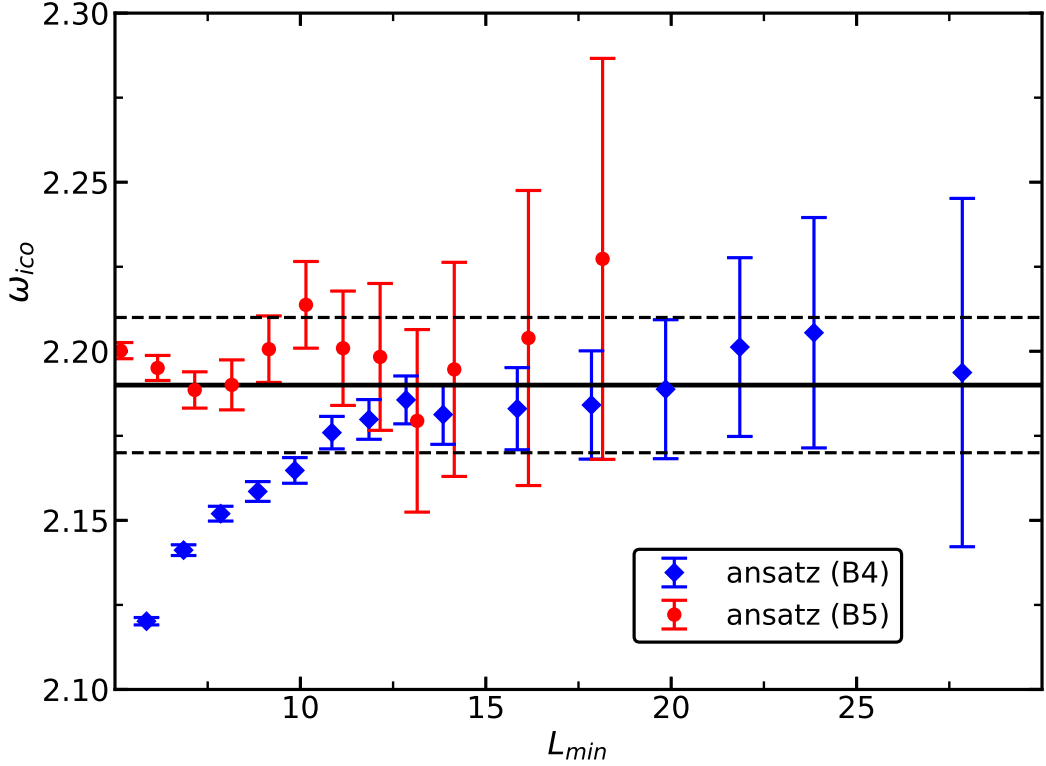}
\caption{\label{omegaicoFig}
We plot the estimate of $\omega_{ico}$ obtained from fitting
$\bar{q}$ at $Z_a/Z_p=0.19477$ as a function of the minimal lattice $L_{min}$ 
size that is taken into account. Data for $D=2.1$ and $2.05$ are jointly
analyzed by using the ans\"atze~(\ref{icofit1},\ref{icofit2}). For readability
the values of $L_{min}$ are slightly shifted. The solid line gives our
final estimate $\omega_{ico}=2.19$, while the dashed lines indicate the 
error.
}
\end{center}
\end{figure}

In Fig. \ref{omegaicoFig} we plot the result of joint fits for $D=2.05$ and
$2.1$ of $\bar{q}$ at $Z_a/Z_p=0.19477$. The free parameters are $a(D=2.05)$,
$a(D=2.1)$, and $\omega_{ico}$ for ansatz~(\ref{icofit1}). In the case of
ansatz~(\ref{icofit2}) $c$ is an additional free parameter, where we assume
$c$ to be the same for $D=2.05$ and $2.1$. 

As our final estimate we take 
\begin{equation}
\omega_{ico} =2.19(2) \;.
\end{equation}
This estimate is chosen such that the estimates obtained by using 
the ansatz~(\ref{icofit1}) for $5 \le L_{min} \le 9$ and the 
ansatz~(\ref{icofit2}) for 
$11 \le L_{min} \le 20$ including the respective error bars are covered.

\end{document}